\documentclass[12pt]{JHEP3}
\usepackage{nicefrac}
\usepackage{graphicx}
\title{Spectral Flow in AdS$_3/$CFT$_2$ }
\preprint{ YITP-SB-07-38}

\author{Gast\'on Giribet\footnote{Email:
gaston@df.uba.ar}$^{~1,2}$,
Ari Pakman\footnote{Email: ari.pakman@stonybrook.edu}$^{~3}$
and Leonardo Rastelli\footnote{Email: leonardo.rastelli@stonybrook.edu}$^{~3}$
\\ \\ \\
\it $^1$ Center for Cosmology and Particle Physics,\\
New York University,  \\
4 Washington Place, NY, 10003,  USA
\\
\\
\it $^2$
Consejo Nacional de Investigaciones Cient\'ificas y T\'ecnicas,\\
Rivadavia 1917, C1033AAJ, Buenos Aires, Argentina
\\
\\
\it $^3$ C.N. Yang Institute for Theoretical Physics,\\
\it Stony Brook University, \\
\it Stony Brook, NY 11794-3840, USA}

\abstract{
We study
 the spectral flowed sectors of the \h3\ WZW model
in the context of the holographic duality between type IIB string theory in $AdS_3 \times S^3 \times T^4$
with NSNS flux and the symmetric product orbifold of $T^4$. We  construct explicitly the physical
 vertex operators in the flowed sectors that belong to short representations of the superalgebra,
 thus  completing the bulk-to-boundary dictionary for 1/2 BPS states. We perform a partial calculation
 of the string three-point functions of these operators. A complete calculation
 would require the three-point couplings of non-extremal flowed operators in the \h3 WZW model,
 which are at present unavailable.
 In the unflowed sector, perfect agreement
 has recently been found between the bulk and boundary three-point functions
 of 1/2 BPS operators. Assuming that this agreement persists in the flowed sectors,
 we determine certain unknown three-point couplings
  in the \h3 WZW model in terms of  three-point couplings
 of affine descendants in the $SU(2)$ WZW model.  }

\newcommand{\sym}{{Sym$^N(M^4)$}}
\newcommand{\symT}{{Sym$^N(T^4)$}}

\newcommand{\nn}{\nonumber}

\def\d{\partial}
\newcommand{\rb}[1]{\raisebox{3.1ex}[0pt]{#1}}
\newcommand{\rbr}[1]{\raisebox{1.5ex}[0pt]{#1}}

\def\a{\alpha}

\def\e{\epsilon}

\def\h{\eta}

\def\up{{\Upsilon}}

\def\O{{\cal O}}

\def\IC{\relax\hbox{$\inbar\kern-.3em{\rm C}$}}

\def\bM{{\bf M}}

\def\IC{{\bf C}}

\def\bea{\begin{eqnarray}}
\def\eea{\end{eqnarray}}
\def\be{\begin{equation}}
\def\ee{\end{equation}}
\def\ea{\end{align}}
\def\bse{\begin{subequations}}
\def\ese{\end{subequations}}
\def\1F1{{}_1\!F_1}
\def\2F0{{}_2\!F_0}

\def\slr{$SL(2,R)$}
\def\ve{\epsilon}

\def\ni{\noindent}
\def\ga{\gamma}

\def\bm{\bar{m}}
\def\bn{\bar{n}}
\def\bx{\bar{x}}
\def\by{\bar{y}}

\def\bH{\bar{H}}
\def\sl{$SL(2,R)$}

\def\nn{\nonumber}

\def\a{\alpha}

\def\h3{$H_3^+$}

\def\d{{\partial}}

\def\IC{{\mathbb C}}

\def\tp{{\tilde \Phi}}

\def\lbldef#1#2{\expandafter\gdef\csname #1\endcsname {#2}}

\def\href#1#2{#2}

\newcommand{\beq}{\begin{equation}}
\newcommand{\eeq}{\end{equation}}
\newcommand{\ber}{\begin{eqnarray}}
\newcommand{\eer}{\end{eqnarray}}

\def\be{\begin{eqnarray}}
\def\ee{\end{eqnarray}}

\def\tj{\tilde{J}}
\def\tjl{\tilde{\jmath}}
\def\tp{\tilde{\psi}}
\def\tk{\tilde{K}}
\def\tkl{\tilde{k}}
\def\tc{\tilde{\chi}}
\def\tz{\tilde{0}}
\def\tV{\tilde{V}}
\def\tP{\tilde{\Phi}}

\def\tn{\tilde{n}}
\def\tm{\tilde{m}}
\def\tT{\tilde{F}}
\def\vac{|0\rangle}

\def\mo{{\mathbb{O}}}
\def\mv{{\mathbb{V}}}
\def\bM{{\bar{M}}}
\def\bJ{{\bar{J}}}

\def\ve{\epsilon}

\def\ni{\noindent}
\def\ga{\gamma}

\def\bm{\bar{m}}
\def\bx{\bar{x}}
\def\by{\bar{y}}

\def\bH{\bar{H}}
\def\sl{$SL(2,R)$}

\def\nn{\nonumber}

\def\a{\alpha}

\def\eb{{\bar{\epsilon}}}

\def\<{\langle}
\def\>{\rangle}

\def\bm{\bar{m}}
\def\bx{\bar{x}}
\def\by{\bar{y}}

\def\bH{\bar{H}}

\keywords{AdS/CFT}

\begin{document}

\section{Introduction}

A classic example of the AdS/CFT correspondence is the duality between type IIB string theory on $AdS_3 \times S^3 \times M^4$,
where $M^4$ is hyperk\"ahler,
and a certain deformation of \sym,
the symmetric product orbifold
of $N$ copies of $M^4$~\cite{Maldacena:1997re}.
The duality can be motivated from  the near-horizon limit of a system of $Q_5$ D5-branes and $Q_1$ D1-branes,
or, in the S-dual frame, of $Q_5$ NS5 branes and $Q_1$ fundamental strings.
The number $N$ of copies of $M^4$ entering  the symmetric product is given\footnote{From now we assume $M^4 = T^4$.}
 by $N = Q_1 Q_5$.

While several aspects of this duality were understood early on
(see {\it e.g.} \cite{Aharony:1999ti, Dijkgraaf:2000vr,David:2002wn, Martinec} for reviews),
only this year has the status of correlation functions become more clear.
Three-point functions of 1/2 BPS operators have been obtained on the string side
 through exact worldsheet computations~\cite{Gaberdiel:2007vu, Dabholkar:2007ey,Pakman:2007hn},
and found to be in precise agreement with the  boundary results of~\cite{Jevicki:1998bm, Lunin:2000yv, Lunin:2001pw}.
Previous supergravity computations~\cite{Mihailescu:1999cj,Arutyunov:2000by,Kanitscheider:2006zf},
which appeared to show a discrepancy, have then been  revisited  \cite{Taylor:2007hs} and found to be compatible with this perfect agreement once a more general ansatz for the bulk-to-boundary dictionary
is assumed.

The string theory and the boundary computations are performed at
different points in the moduli space  \cite{Dijkgraaf:1998gf,Larsen:1999uk},
where solvable descriptions are available. This strongly suggests the existence of a new non-renormalization theorem.
In the boundary, the solvable point is the orbifold \sym.
In the bulk, it is near horizon geometry of the NS5 F1 system, which is
$AdS_3 \times S^3 \times M^4$ with only NSNS flux in the $AdS_3 \times S^3$ factors~\cite{Maldacena:1998bw}.
This leads  to an exact worldsheet description in the RNS formalism
in terms of  $SL(2,R)_{k+2}$ and $SU(2)_{k-2}$ current algebras, where $k \equiv Q_5$, plus some free fermions \cite{Giveon:1998ns, Kutasov:1999xu}.
In this setting the superconformal invariance of the dual theory
can be seen to arise from the string worldsheet~\cite{Giveon:1998ns,deBoer:1998pp, Giveon:2003ku}.

Let us recall the basics of the bulk-to-boundary dictionary for 1/2 BPS operators.
The six-dimensional string coupling constant is
\be
g_6^2= \frac{Q_5}{Q_1}\,,
\ee
 so string perturbation theory is valid for $Q_1 \rightarrow \infty$, or $N = Q_1 Q_5 \gg Q_5$.
In this limit, single string states in the bulk map to twisted
states in \sym\ associated to conjugacy classes with a single
non-trivial cycle. The length $n$ of this cycle is related to the
$SL(2,R)$ spin $h$ appearing in the worldsheet vertex operator as
\be
n= 2h-1\, . \label{nh}
\ee
The analysis of~\cite{Gaberdiel:2007vu,
Dabholkar:2007ey,Pakman:2007hn} included only operators arising
from the usual ``unflowed'' representations of the $SL(2,R)$ current
algebra, whose spin is bounded as $h \leq k/2$. On the other hand,
the symmetric product orbifold contains  cycles of lengh $n \leq N
$. So in the large $N$ limit, where the worldsheet description is
valid, it appears that infinitely many 1/2 BPS operators are missing
in the bulk.  The resolution of this puzzle has been known for some
time: the additional operators arise
\cite{Hikida:2000ry,Argurio:2000tb} from the spectral flowed sectors
of the \sl\ current algebra~\cite{Maldacena:2000hw,
Maldacena:2000kv, Maldacena:2001km}.
Once spectral flowed representations are included, relation~(\ref{nh})
is generalized to
\be
n= 2h-1 +k w \,,  \qquad \qquad h = 1\,, \frac{3}{2} \, ,  \ldots \frac{k}{2} \, , \qquad w =0\, ,1 \, , 2 \, ,\ldots
\label{nhw}
\ee
where $w$ is the spectral flow parameter.\footnote{As noticed in \cite{Argurio:2000tb}, the range of $n$ in (\ref{nhw}) is such that the values $n=kw$ are still absent.
The singular nature of the boundary CFT \cite{Seiberg:1999xz}
may be related to this fact. See also the recent discussion in \cite{Raju:2007uj}. As described in \cite{Rastelli:2005ph}, there is a connection
between the $AdS_3 \times S^3$ background and the minimal $(k,1)$ string. In the minimal string, the absence of the corresponding states is natural
from the viewpoint of the KP integrable hierarchy. Another curious observation \cite{Rastelli:2005ph} is that the first missing state
($w=1$) has the quantum numbers of an {\it  open} string state on the $H_2^+$ brane. This
is again similar to  the situation in the minimal string \cite{Martinec:1991ht}.}

In this paper we give a precise construction of the 1/2 BPS vertex operators
in the flowed sectors, thus completing the bulk-to-boundary dictionary, and we study their three-point functions.
The physical states in the flowed sectors are the $AdS_3$ analogs of what in flat space are the infinite
higher-spin string modes -- they are genuine string states not
visible in supergravity. Indeed, supergravity becomes a good description for $k \to \infty$,
and in this limit the  flowed states acquire infinite conformal dimension.
The BPS condition
correlates the $SL(2,R)$ and $SU(2)$ quantum numbers,
and the complete vertex operators involve a precise combination of
 states from the~$SU(2)$ WZW model and  the  worldsheet fermions,
which are current algebra descendants but Virasoro primaries.

As in \cite{ Dabholkar:2007ey},
we obtain the vertex operators in the ``$x$-basis'', which greatly facilitates explicit
computations. We then perform a partial calculation of three-point functions of flowed operators.
Some features of the calculation
 suggest that the agreement with the boundary results continues to hold in the flowed sectors.
 Unfortunately, a complete calculation requires
 certain three-point couplings in the \h3 WZW model that are not yet available
  in the CFT literature.  We leave their evaluation
 for future work. Instead of a complete verification of the bulk-to-boundary agreement,
we turn the logic around and obtain non-trivial holographic predictions in the form of identities involving
three-point couplings of flowed operators of the \h3\ WZW model and of affine descendants of~the~$SU(2)$~WZW model.

The organization of the paper is as follows.
In Section \ref{symprod} we review  the spectrum and
three-point functions of 1/2 BPS operators in the boundary theory, and
the spectrum and three-point functions of {\it unflowed} 1/2 BPS operators in the
the bulk theory. In Section \ref{spectral}, we review the spectral flow in the $SL(2,R)$ affine algebra and
study its $SU(2)$ counterpart,  which maps current algebra primaries to descendants.
In Section \ref{sfff}, we study how the spectral flow organizes the spectrum of the free fermions
into $SL(2,R)$ and $SU(2)$ multiplets of Virasoro primaries, and compute their three-point functions.
In Section \ref{chiral-spectrum} we assemble our previous results to build the flowed 1/2
BPS physical vertex operators. In Section \ref{chiral-3pf} we study their three-point functions and
 obtain the identities that must hold assuming the bulk-to-boundary agreement.  We conclude in
Section 7.

\section{Review of 1/2 BPS operators and  their three-point functions \label{symprod}}

\subsection{The symmetric product orbifold }

In this subsection we briefly review the ${\cal N}=2$ and ${\cal N}=4$ spectrum and three-point correlators of the
symmetric product orbifold \sym. For more details see \cite{Vafa:1994tf, Jevicki:1998bm, Lunin:2000yv, Lunin:2001pw}.

There is one twisted sector for each conjugacy class of the symmetric group $S_N$,
given by disjoint cycles of lengths $n_i$ and multiplicities $N_i$,
\be
(n_1)^{N_1}(n_2)^{N_2} \ldots (n_r)^{N_r} \, , \qquad
\sum _i n_i N_i = N \,.
\ee
According to the AdS/CFT dictionary, in the large $N$ limit each cycle is interpreted as a single string state.
Therefore, the chiral primary operators that we are interested in
are given by twist fields
associated to single cycles of length $n_i$,
dressed by chiral fields of $M^4$ itself, summed  in a~$S_N$ invariant way.
For $M^4=T^4$, the holomorphic chiral fields are $1$, $\psi^a \, (a=1,2)$ and $\psi^1\psi^2$,
where~$\psi^1$ and~$\psi^2$ are complex fermions formed by grouping the four fermions of $T^4$ into two pairs.
We will label the chiral operators dressed by $1$, $\psi^a$ and $\psi^1\psi^2$ respectively as\footnote{The operators of type $-$ and $+$ were
called type $0$ and $2$ in \cite{Jevicki:1998bm, Dabholkar:2007ey}.}   $O_{n}^{-}, O_{n}^{a}$ and $O_{n}^{+}$.
Their holomorphic conformal dimensions are respectively $\Delta = \frac{n-1}{2},\frac{n}{2}$ and $\frac{n+1}{2}$.
Each operator has also an independent anti-holomorphic dressing, so the full operators
are denoted as $O_{n}^{(\e,\eb)}$, where $\e, \eb = -, a, +$.

These operators are ${\cal N} =2$ chiral, namely their conformal dimension $\Delta$ and $U(1)$ R-charge $Q$ satisfy~$\Delta=Q/2$.
Since \sym\ has ${\cal N}=4$ superconformal symmetry, with  affine left and right $SU(2)_{N}$ R-symmetries, these
are actually the highest weight states in $SU(2) \times SU(2)$ multiplets of spins $J=\Delta$ and $\bar{J}= \bar{\Delta}$.
Denoting the elements of these multiplets as
$\mv_{n,M,\bM}^{(\e, \eb)}$,
 we have
\be
O_n^{(\e, \eb)} &=& \mv_{n,J,\bJ}^{(\e, \eb)} \,, \\
O_n^{(\e, \eb) \,\dagger} &=& \mv_{n,-J,-\bJ}^{(\e, \eb)} \, , \qquad \e, \eb= -, a,+ \,\,.
\ee
It is convenient to normalize the modes $\mv_{n,M,\bM}^{(\e, \eb)}$  as
\be
\< \mv_{n,M,\bM}^{(\e,\eb)} \mv_{n',M',\bM'}^{(\e', \eb')} \rangle = (-1)^{J + \bJ - M - \bM} \delta_{n n'} \delta^{\e\e'} \delta^{\eb\eb'}  \delta_{M M'} \delta_{\bM \bM'}  \,.
\ee
We then sum  over formal isospin variables $y,\by$ to define
\be
 \mo_n^{(\e,\eb)} (y,\by) =  \sum_{M=-J}^{J} \sum_{\bM=-\bJ}^{\bJ}
 \left( c_{M}^{J} c_{\bM}^{\bJ} \right)^{\nicefrac{1}{2}}
 \times   y^{-M+J} \bar{y}^{-\!\bM+\bJ} \mv_{n,M,\bM}^{(\e,\eb )}
\label{vm}
\ee
where
\be
c_{M}^{J} = \left(\begin{array}{c} 2J \\ M+J \end{array}\right) = \frac{(2J)!}{(J+M)!(J-M)!} \,.
\ee
The operators $\mo_n^{(\e,\eb)}$ obey
\be
\langle
\mo_n^{(\e,\eb)} \mathbb{O}_{n'}^{(\e',\eb')} \rangle
= (y_1-y_2)^{2J} (\bar{y}_1 - \bar{y}_2)^{2\bar{J}} \delta_{n n'} \delta^{\e\e'} \delta^{\eb\eb'} \, .
\label{on2pf}
\ee
In summary, there are three series of holomorphic $SU(2)$ multiplets
$\mo_{n}^{(-)}(y), \mo_{n}^{(a)}(y)$ and $\mo_{n}^{(+)}(y)$,
with $\Delta=J= \frac{n-1}{2}, \frac{n}{2}$ and $\frac{n+1}{2}$, respectively, where $n=1,\ldots, N$.
Similarly, there are three series of antiholomorphic multiplets that depend on the $\by$ isospin variable.
The full 1/2 BPS spectrum is obtained by putting together holomorphic and antiholomorphic multiplets,
with the constraint that cycle length $n$ be the same for  the holomorphic and antiholomorphic
factors.

\subsection*{Extremal $N=2$ correlators}

The three-point functions for the ${\cal N}=2$ primaries $O_n^{(\e, \eb)}$, which  correspond to  ``extremal'' correlators in the terminology of \cite{D'Hoker:1999ea},
were computed in \cite{Jevicki:1998bm}. The fusion rules, obtained from conservation of
the~$U(1)$ R-charge and
the group composition law of the cyclic permutations, are,
\be
\begin{array}{rclclcr}
(-) & \times & (-) &\rightarrow& (-) &+& (+)  \\
(-) & \times & (+) &\rightarrow& (+) & &          \label{fusionrules} \\
(-) & \times & (\,a\,) &\rightarrow& (\,a\,) & &      \\
(\,a\,) & \times & (\,a\,) &\rightarrow& (+) & &
\end{array}
\ee
both for the holomorphic and the anti-holomorphic sectors.
These fusion rules are combined  freely between both sectors, except for the process
\be
(-)  \times  (-) \rightarrow (+)
\ee
which should occur simultaneously in the left and the right movers.
This gives a total combination of $4 \times 4 + 1 =17$ possible fusions.
In the large $N$ limit, the structure constants for the scalar~$(\Delta= \bar{\Delta})$ sector are
\be
\label{boundary3pf1}
\langle O_{n_3}^{(-,-) \,\dagger }  O_{n_2}^{(-,-)} O_{n_1}^{(-,-)} \rangle
&=& \left(\frac{1}{N}\right)^{1/2} \, \left(\frac{n_3 ^3}{n_1 n_2} \right)^{1/2}
\\
\label{boundary3pf2}
\langle O_{n_3}^{(+,+) \,\dagger}  O_{n_2}^{(-,-)} O_{n_1}^{(-,-)} \rangle
&=&
 \left(\frac{1}{N}\right)^{1/2}
\left(\frac{1}{n_1 n_2 n_3} \right)^{1/2}
\\
\langle O_{n_3}^{(+,+)\,\dagger}  O_{n_2}^{(-,-)} O_{n_1}^{(+,+)} \rangle
&=&
\left(\frac{1}{N}\right)^{1/2}
\left(\frac{n_1^3}{n_2n_3} \right)^{1/2}
\\
\label{boundary3pf4}
\langle O_{n_3}^{(\, a,\bar{a}\, ) \, \dagger}  O_{n_2}^{(\, b,\bar{b} \,)} O_{n_1}^{(-,-)} \rangle
&=&
\left(\frac{1}{N}\right)^{1/2}
\left(\frac{ n_2 n_3}{n_1} \right)^{1/2} \delta^{ab} \delta^{\bar{a}\bar{b}}
\\
\label{boundary3pf5}
\langle O_{n_3}^{(+,+) \, \dagger}  O_{n_2}^{(\,a,\bar{a}\,)} O_{n_1}^{(\, b,\bar{b}\,)} \rangle
&=&
\left(\frac{1}{N}\right)^{1/2}
\left(\frac{ n_1 n_2}{n_3} \right)^{1/2} \xi^{a\,b}  \xi^{\bar{a}\,\bar{b}} \,,
\ee
with
\be
\xi &=&
\left(
\begin{array}{rr}
0 & 1 \\
1 & 0
\end{array}
\right) \,.
\ee
Here $n_3$ is fixed in terms of $n_1$ and $n_2$ from the conservation of $U(1)$ R-charge,
which gives ~$n_3=n_1+n_2 -3$ for~(\ref{boundary3pf2}) and by~$n_3=n_1+n_2-1$ for the other  cases.
The structure constants are actually completely factorized between left and right movers,
so for non-scalar operators the three-point functions are products of  square roots of the above correlators.

\subsection*{Non-extremal $N=4$ correlators}

Correlators involving the elements of the full $SU(2)$ multiplet were computed, in \cite{Lunin:2000yv,Lunin:2001pw},
only for operators of type $\e,\eb=\pm$. The fusion rules are
\be
&& n_i \leq n_j + n_k -1 \,, \qquad \qquad i\neq j\neq k \qquad i,j,k=1,2,3 \,.
\label{boundary-fus}
\ee
Their three-point functions are, in the large $N$ limit,
\be
&\<\mv_{n_1,M_1,\bM_1}^{(\e_1,\eb_1)}\mv_{n_2,M_2,\bM_2}^{(\e_2,\eb_2)}\mv_{n_3,M_3,\bM_3}^{(\e_3,\eb_3)}\>
=
\frac{1}{\sqrt{N}}
\frac{(\e_1 n_1 + \e_2n_2 + \e_3n_3 +1 ) (\eb_1 n_1 + \eb_2n_2 + \eb_3n_3 +1 ) }
{4  (n_1 n_2 n_3)^{\nicefrac12}}
\\
\nn
& \times L(J_i,M_i) L(\bJ_i,\bM_i) \, \delta_{M_1 + M_2 + M_3, 0}  \delta_{\bM_1 + \bM_2 + \bM_3, 0}  \, ,
\ee
where $L(J_i,M_i)$ is defined in terms of the $SU(2)$  $3j$ symbols as
\be
& L(J_i,M_i) = \left( \begin{array}{ccc} J_1&J_2&J_3 \\M_1&M_2&M_3 \end{array}\right)
 \times  \left[ \frac{(J_1+J_2-J_3)! (J_2+J_3-J_1)! (J_3+J_1-J_2)! (J_1+J_2+J_3+1)!}{(2J_1)!(2J_2)!(2J_3)!}\right]^{\nicefrac12} \,.
\ee
In terms of the $\mo_{n}^{(\e,\eb)}$ multiplets defined in (\ref{vm}), the three-point functions take  the simple form~\cite{Pakman:2007hn}
\begin{eqnarray}
\label{sym}
&\langle \mo_{n_1}^{(\e_1,\eb_1)}  \mo_{n_2}^{(\e_2,\eb_2)}
\mo_{n_3}^{(\e_3,\eb_3)}   \rangle = \frac{1}{\sqrt{N}}\frac{(\e_1 n_1 + \e_2n_2 + \e_3n_3 +1 ) (\eb_1 n_1 + \eb_2n_2 + \eb_3n_3 +1 ) }
{4(n_1 n_2 n_3)^{\nicefrac12}}
\\
 \nn
& \times \,\,
y_{12}^{J_1+J_2-J_3}
y_{23}^{J_2+J_3-J_1}
y_{31}^{J_3+J_1-J_2}  \times \,\,
\by_{12}^{\bar{J}_1+\bar{J}_2-\bar{J}_3}
\by_{23}^{\bar{J}_2+\bar{J}_3-\bar{J}_1}
\by_{31}^{\bar{J}_3+\bar{J}_1-\bar{J}_2} \,.
 \label{PD}
\end{eqnarray}
One can easily verify that these $N=4$ correlators reduce to the extremal $N=2$ correlators
when we specialize to $M_i=\pm J_i$ and $\bM_i=\pm \bar{J}_i$.

\subsection{The $AdS_3 \times S^3 \times T^4$ worldsheet}
In the frame with only NSNS flux, the string background
 is described by a product of supersymmetric \sl\ and $SU(2)$ WZW models at level~$k$,
which correspond to the~$AdS_3 \times S^3$ geometry~\cite{Maldacena:1998bw, Giveon:1998ns, Kutasov:1999xu}, and four real bosons and fermions,
corresponding to the $T^4$ factor. We will actually consider the Euclidean form
of $AdS_3$, where the \sl\ WZW model is replaced by the \h3\ WZW model, and whose affine
symmetries are still two copies of \sl.

The supersymmetric affine $SL(2,R)_k$  symmetry is
generated by the supercurrents $\psi^A + \theta J^A$, $A=1,2,3$.
The OPEs are
\be
J^A(z) J^B(w) \sim &&
{{k\over 2} \eta^{AB} \over (z-w)^2} + {i\epsilon^{AB} {}_{C} J^C(w) \over z-w}~,
\label{jjope}
\\
J^A(z) \psi^B(w) \sim && {i\epsilon^{AB}{}_{C} \psi^C(w)
\over z-w}~,
\label{jpsi}
\\
\psi^A(z) \psi^B(w) \sim && { {k\over 2}\eta^{AB} \over z-w}~,
\ee
where $\epsilon^{123}=1$ and capital letter indices are raised and
lowered with $\eta^{AB}=\eta_{AB}=(++-)$. Similarly, the
supersymmetric affine $SU(2)_k$ symmetry has supercurrents $\chi^a + \theta
K^a$, $a=1,2,3$, with OPEs \be K^a(z) K^b(w) \sim && {{k\over 2}
\delta^{ab} \over (z-w)^2} + {i\epsilon^{ab}{}_{c}
 K^c(w) \over z-w}~,
 \label{kkope}
 \\
K^a(z) \chi^b(w) \sim && {i\epsilon^{ab}{}_{c}  \chi^c(w)
\over z-w}~,\\
\chi^a(z) \chi^b(w) \sim && { {k\over 2} \delta^{ab} \over z-w}~,
\ee
and lower case indices are raised and lowered with $\delta^{ab}=\delta_{ab}=(+,+,+)$.
We will often use the linear combinations
\be
J^{\pm} &\equiv& J^1 \pm i J^2 \qquad \qquad \psi^{\pm} \equiv \psi^1 \pm i \psi^2 \,, \\
K^{\pm} &\equiv& K^1 \pm i K^2 \qquad \qquad \!\!\! \chi^{\pm} \equiv \chi^1 \pm i \chi^2 \,.
\ee
As usual in supersymmetric WZW models,
it is convenient to split the $J^A, K^a$  currents into
\be
J^A &=& j^A +   \hat{ \jmath}^A \,,
\label{jsplit}
\\
K^a &=& k^a +\hat{k}^a\,,
\label{ksplit}
\ee
where
\be
\hat{\jmath}^A &=& -\frac{i}{k} \epsilon^{A}{}_{BC} \psi^B \psi^C \,, \\
\hat{k}^a &=& -\frac{i}{k} \epsilon^{a}{}_{bc} \chi^b \chi^c \,.
\ee
The currents $j^A$ and $k^a$ generate  bosonic  $SL(2,R)_{k+2}$ and
$SU(2)_{k-2}$  affine algebras,
and commute with the free fermions $\psi^A,\chi^a$.  The latter in turn
form a pair of supersymmetric \slr\ and $SU(2)$ models at levels -2
and +2, whose bosonic currents are $\hat{\jmath}^A$ and~$\hat{k}^a$.
The spectrum and the  interactions of the original level $k$
supersymmetric WZW models are factorized into the bosonic WZW models
and the free fermions~\cite{Fuchs:1988gm}.
In terms of the split currents, the stress
tensor and supercurrent of $SL(2,R)$ are
\be
T^H &=&  \frac{1}{k}j^Aj_A - \frac{1}{k} \psi^A \d \psi_A \,,
\\
G^H &=&   \frac{2}{k} ( \psi^A j_A + \frac{2i}{k} \psi^1\psi^2\psi^3)
\,,
\ee
and those of $SU(2)$ are
\be
T^S &=&  \frac{1}{k}k^ak_a  - \frac{1}{k} \chi^a \d \chi_a    \,, \\
G^S &=&  \frac{2}{k}(\chi^a k_a - \frac{2i}{k}\chi^1\chi^2\chi^3) \, .
\ee
The total stress tensor and supercurrent are
\be
T &=& T^H + T^S + T(T^4) \\
G &=& G^H + G^S + G(T^4)
\ee
where $T(T^4)$ and $G(T^4)$ are the
stress tensor and supercurrent of $T^4$,
and one can check that the central charge adds up to $c=15$.
Here and below we focus on the holomorphic part of the theory, but there is a similar
antiholomorphic copy.

A primary field of spin $h$ in the $SL(2,R)_{k+2}$ WZW model satisfies
\be
j^A(z)
\Phi_{h}(x,w) \sim - \frac{D_x^A \Phi_{h}(x,\bx;w,\bar{w})}{z-w} \,,
\label{jphi}
\ee
where the operators $D_x^A$ are
\be
D_x^- &=& \d_x \,,
\label{dm}
\\
D_x^3 &=& x \d_x + h \,,
\label{d3}
\\
D_x^+ &=& x^2 \d_x + 2hx \,.
\label{dp}
\ee
The conformal dimension of $\Phi_h$ is
\be
\Delta_h=  -\frac{h(h-1)}{k}\,.
\ee
The field $\Phi_h$  can be expanded in modes as
\be
\Phi_h(x,\bx) = \sum_{m,\bm} \Phi_{h,m,\bm} x^{-h-m} \bx^{-h-\bm}\,,
\ee
but the range of the summation is not always
well defined \cite{Kutasov:1999xu}. Yet, the action of the zero
modes of the currents on $\Phi_{h,m,\bm}$  is well defined and can
be read from (\ref{jphi}) to be
\be
j^3_0 \Phi_{h,m,\bm} &=& m
\Phi_{h,m,\bm} \,\,\,\,,
\\
j^{\pm}_0 \Phi_{h,m,\bm} &=& (m \mp (h-1) ) \Phi_{h,m\pm1,\bm}
\,\,\,\,,
\ee
and similarly for the anti-holomorphic currents.
The $x,\bx$  variables
are interpreted as the local coordinates of the two-dimensional
conformal field theory living in the boundary of $AdS_3$.

The Hilbert space of the $SL(2,R)_{k+2}$ consists of
 the usual ``unflowed'' sector and of the spectral flowed sectors \cite{Maldacena:2000hw, Maldacena:2000kv}.
Let us recall the structure of the unflowed sector. As usual,
we can decompose it in representations of the current algebra
built by the action of the negative modes of $j^A$ on affine primaries.
In  turn, the affine primaries form representations of the $SL(2,R)$ algebra of the zero modes.
The relevant representations of the zero modes are
delta-normalizable continuous
representations, with $h=\frac12 + i \mathbb{R}$ and $m= \a +
\mathbb{Z} \, (\a \in [0,1))$, and non-normalizable discrete
representations, with~$h \in \mathbb{R}$ obeying
\be
\frac12 < h <
\frac{k+1}{2}\,.
\label{hboundf}
\ee
The discrete representations can be either lowest-weight $d^+_h$,
with $m=h,h+1\ldots$, or highest-weight $d^-_h$, with $m=-h,-h-1 \ldots$.
The  spectral flowed sectors of the Hilbert space will be considered in the next section.

The bosonic $SU(2)_{k-2}$ WZW model has primaries $V_{j,m,\bm}$ with
$m,\bm=-j,\ldots,+j$, and the spin $j$ is bounded by \cite{Zamolodchikov:1986bd,Gepner:1986wi}
\be
0 \leq j \leq \frac{k-2}{2} \,.
\label{jboundf}
\ee
The conformal dimension of $V_{j,m,\bm}$ is
\be
\Delta = \frac{j(j+1)}{k}\,.
\ee
Similarly to the $x,\bx$ variables of the
\slr\ sector,  isospin coordinates $y,\by$ can be introduced for
$SU(2)$ \cite{Zamolodchikov:1986bd},  such that the primaries are organized
into the fields $V_j$,
\be
V_j(y,\by) \equiv \sum_{m=-j}^{j} V_{j,m,\bm} y^{-m+j} \by^{-\bm
+j} \,\,\,.
\ee
The action of the $k^a$ currents on $V_j(y;z)$ is
\be
k^a(z) V_{j}(y;w) \sim - \frac{P_y^a V_j(y;w)}{z-w}  \,,
\label{kv}
\ee
where the differential operators
\be
P_y^- &=& -\d_y
\label{pm}
\\
P_y^3 &=& y \d_y - j \\
P_y^+ &=& y^2 \d_y - 2jy
\label{pp}
\ee
are the $SU(2)$ counterparts of
$D_{x}^A$. There is a  similar antiholomorphic copy. The  action
of the zero modes of $k^a$  on $V_{j,m,\bm}$ can be read from
(\ref{kv}) to be
\be k^3_0 V_{j,m,\bm} &=& m V_{j,m,\bm}
\label{k3action}
\\
k^{\pm}_0 V_{j,m,\bm} &=& (\pm m +1 +j ) V_{j,m\pm1,\bm} \,\,\,\,\, (m\neq \pm j)
\label{kpmaction}
\\
k^{+}_0V_{j,j,\bm} &=&
k^{-}_0V_{j,-j,\bm} = 0 \, ,
 \ee
and similarly for $\bar{k}^a_0$.

The 1/2 BPS vertex operators in the bulk that correspond to the boundary operators $\mo_{n}^{(\e,\eb)}$
are $SU(2)$ multiplets obeying
\be
H=J \qquad \bar{H}=\bar{J},
\ee
where the upper-case spins $H$ and $J$ are similar to  $h$ and $j$
but measured with respect to the full algebras $J^A$ and $K^a$.
In the unflowed sector of $SL(2,R)$ these chiral states were obtained in \cite{Kutasov:1998zh} in the $m,n$ basis,
and were recast  in $x, y$ basis in \cite{Dabholkar:2007ey}.
One finds that both in the holomorphic and anti-holomorphic sectors
there are three families of operators,
in 1-1 correspondence with the operators $\mo_{n}^{(-)}, \mo_{n}^{(a)}$ and $\mo_{n}^{(+)}$
of the symmetric orbifold.
Basic building blocks are the $k-1$ affine primaries
\be
{\cal O}_h(x,y) \equiv  \Phi_{h} (x) V_{h-1}(y) \qquad \qquad \qquad h = 1,\frac32, \ldots \frac{k}{2} \,,
\ee
which have  $\Delta (\O_h(x,y))=0$.
In the holomorphic sector, in the $-1$ ($-1/2$) picture of the NS (R) sector,
the three families of operators are given by\footnote{The operators of type $-$ and $+$ were called type $0$ and $2$ in \cite{Dabholkar:2007ey}.}
\be \label{unflowedfamilies}
\O_h^{(-)}(x,y)  &= & e^{-\phi} \O_h(x,y) \psi(x) \qquad \qquad H=J=h-1
\label{obm}
\\
\O_h^{(+)}(x,y)  &= & e^{-\phi} \O_h(x,y) \chi(y) \qquad \qquad H=J = h\\
\O_h^{(a)}(x,y)  &= & e^{-\frac{\phi}{2}} \O_h(x,y) s^{a}_-(x,y)\qquad   \,\, H=J = h-\frac{1}{2}   \qquad \qquad a=1,2
\label{oba}
\ee
where
\be
\psi(x) &=& -\psi^+ +2x \psi^3 -x^2\psi^-\,, \label{psidef}
 \\
\chi(y) &=& -\chi^+ +2y \chi^3 +y^2\chi^- \,, \label{chidef}
\ee
and $s^{a}_{-}(x,y)$ are spin fields whose explicit form can be found below in (\ref{oa-exp}).
Here  $\phi$ is the usual boson coming from the bosonization of the $\beta \gamma$ ghosts~\cite{Friedan:1985ge}.
The full vertex operators are obtained by dressing the above expressions with the  anti-holomorphic
operators $\bar{\psi}(\bar{x}),\bar{\chi}(\bar{y})$ and $\bar{s}^{a}_{\pm}(\bar{x},\bar{y})$.

After  normalizing the bulk vertex operators as in~(\ref{on2pf}), it was shown
in~\cite{Gaberdiel:2007vu, Dabholkar:2007ey,Pakman:2007hn} that their three-point functions
agree with those of the boundary, under the identification
\be
n = 2h-1 \,.
\ee
The range of $h$ and of the correlated quantum number $j = h-1$ are restricted by
the bounds~(\ref{hboundf}) and~(\ref{jboundf}). We see that there are $k-1$ operators
of each type.

As  explained in the Introduction, in the
symmetric orbifold the quantum number $n$ can be an arbitrary positive integer.
The missing bulk vertex operators arise from the spectral flowed sectors of $SL(2, R)$.
A key point is that while all the operators in (\ref{unflowedfamilies})
are built from affine primaries, BRST invariance is less restrictive and only requires the operators to be (super)Virasoro primaries.
In Section \ref{chiral-spectrum} we will find that
that each family of physical vertex operators admits infinitely many spectral flowed relatives
\be
\O_{h,w}^{(-)}  &= & e^{-\phi} \Phi_{h,w}  V_{h-1,w} \, \psi_{w+1} \, \chi_{w}  \qquad \qquad H=J=h + \frac{kw}{2}-1
\label{op1flow}
\\
\O_{h,w}^{(+)}  &= & e^{-\phi} \Phi_{h,w}  V_{h-1,w} \, \psi_{w} \, \chi_{w+1} \; \; \;\quad \qquad H=J = h + \frac{kw}{2}
\label{op2flow}
\\
\O_{h,w}^{(a)}  &= & e^{-\frac{\phi}{2}} \Phi_{h,w}  V_{h-1,w} \, s^{a}_{w, -}\qquad \quad \qquad \,\,  H=J = h + \frac{kw}{2} -\frac{1}{2}   \, ,
\label{op3flow}
\ee
where  $a=1,2$ and $w$ is a non-negative integer. The bulk-to-boundary dictionary generalizes to
\be
n = 2h-1 + kw \,.
\ee
Here $\Phi_{h,w}$ are operators in the spectral flowed sectors of \h3, $V_{h-1,w}$ are
multiplets of the global $SU(2)$ symmetry built from affine algebra descendants,
and~$\psi_{w}$, $\chi_{w}$ and~$s^{a}_{w, -}$ are $SU(2)$ and \sl\ multiplets that are descendants
in the Hilbert space of the fermions. All these fields are Virasoro primaries.

In Section \ref{spectral} we study in detail the fields
$\Phi_{h,w}$ and $V_{j,w}$. In  Section  \ref{sfff} we consider the fields $\psi_{w}$, $\chi_{w}$ and~$s^{a}_{w, -}$ and their interactions.
In Section \ref{chiral-spectrum} we assemble these ingredients to obtain
the 1/2 BPS vertex operators.

\section{Spectral Flow in $SL(2,R)$ and $SU(2)$ \label{spectral}}

The modes of the $SL(2,R)_k$  currents satisfy
\be
[ J^3_n, J^3_m ]    &=& - \frac{k}{2} n \delta_{n+m,0}
\\
\left[ J^3_n, J^{\pm}_m \right]  &=& \pm J^{\pm}_{n+m}
\\
\left[ J^+_n, J^-_m \right]  &=& -2J^3_{n+m} + kn\delta_{n+m,0}
\\
\left[ J^3_n, \psi^\pm_m \right]&=& \pm \psi^\pm_{n+m}
\\
\left[ J^\pm_n,\psi^\mp_m \right] &=& \mp 2\psi^3_{n+m}
\\
\left[ J^\pm_n, \psi^3_m \right] &=& \mp \psi^\pm_{n+m}
\\
\{ \psi^3_n,\psi^3_m \} &=& -{k \over 2} \delta_{n+m,0}
\\
\{ \psi^+_n,\psi^-_m \}&=& k \delta_{n+m,0} .
\ee
The $SU(2)_k$ modes satisfy
\be
[ K^3_n, K^3_m ]    &=&  \frac{k}{2} n \delta_{n+m,0}
\\
\left[ K^3_n, K^{\pm}_m \right]  &=& \pm K^{\pm}_{n+m}
\\
\left[ K^+_n, K^-_m \right]  &=& 2K^3_{n+m} + kn\delta_{n+m,0}
\\
\left[ K^3_n, \chi^\pm_m \right]&=& \pm \chi^\pm_{n+m}
\\
\left[ K^\pm_n,\chi^\mp_m \right] &=& \pm 2\chi^3_{n+m}
\\
\left[ K^\pm_n, \chi^3_m \right] &=& \mp \chi^\pm_{n+m}
\\
\{ \chi^3_n,\chi^3_m \} &=& {k \over 2} \delta_{n+m,0}
\\
\{ \chi^+_n,\chi^-_m \}&=& k \delta_{n+m,0} .
\ee
Both algebras have spectral flow isomorphisms, corresponding to the replacements
$J^A,\psi^A \rightarrow \tj^A,\tp^A$ and $ K^a, \chi^a \rightarrow  \tk^a, \tc^a$. For $SL(2,R)_k$,
\be
\tj^3_n &=& J^3_n - \frac{k}{2}w\delta_{n,0}
\label{j3h}
\\
\tj^{\pm}_n &=& J^\pm_{n \pm w} \\
\tp^3_n &= & \psi^3_n \\
\tp^{\pm}_n &= & \psi^{\pm}_{n \pm w} \, ,
\ee
where $w$ is an integer. For $SU(2)_k$,
\be
\tk^3_n &=& K^3_n + \frac{k}{2}w\delta_{n,0} \\
\tk^{\pm}_n &=& K^\pm_{n \pm w} \\
\tc^3_n &= & \chi^3_n \\
\tc^{\pm}_n &= & \chi^{\pm}_{n \pm w} \,.
\ee

Let us adopt the collective names
\be
F &=& \{j^A,\psi^A, k^a, \chi^a \} \\
\tT &= & \{ \tilde{\jmath}^A,\tp^A, \tilde{k}^a, \tc^a \}
\ee
The spectral flow is useful because it maps
one representation of the affine algebra into another.
For this, we build a representation with  the unflowed~$\tT$ generators,
and read its quantum numbers in the spectral flowed frame $F$ with   $L_0^H, J^3_0, L_0^S, K^3_0$,
which are given by
\be
J_0^3 &=& \tilde{J}^3_0 + w\frac{k}{2} \\
L^H_0 &=& \tilde{L}^H_0 - w\tilde{J}_0^3 -\frac{k}{4} w^2
\label{lzh}
\ee
and
\be
K_0^3 &=& \tilde{K}^3_0 - w\frac{k}{2}
\label{kz}
\\
L^S_0 &=& \tilde{L}^S_0 - w \tilde{K}_0^3 +\frac{k}{4} w^2 \,.
\label{lzs}
\ee
This map has a very different nature in the $j^A$ sector than in the
 $k^a, \chi^a$ and $\psi^A$ sectors.
For these last cases, the spectral flow amounts to a reshuffling of different representations which maps
primaries to descendants.
On the other hand, in the $j^A$  sector it
generates new representations,
whose~$L_0$ values are unbounded from below (see {\it e.g.} \cite{Feigin:1997ha}).
In the context of strings propagating in $AdS_3$ backgrounds, it was shown by Maldacena and Ooguri
in \cite{Maldacena:2000hw,Maldacena:2000kv,Maldacena:2001km}
that it is necessary to include these new representations in
order to solve several consistency problems, such as an unnatural bound on the
excitations of the inner theory and the identification of long strings~\cite{Seiberg:1999xz,Maldacena:1998uz},
and to obtain a modular invariant partition function~\cite{Maldacena:2000kv,Israel:2003ry}.

\subsection{Spectral Flow in Bosonic \sl}
A general feature of spectral flow, which will be very important for us,
is that an  {\it affine} primary state in~$\tT$, is mapped
to a Virasoro primary in~$F$, which is moreover a highest/lowest weight
state in a $d^{\pm}$ representation
of the {\it global} algebra~\cite{Maldacena:2001km}.
Indeed, consider a highest weight state~$|\varphi\rangle$ of the
affine algebra~$\tjl^A_s$, satisfying
\be
\tjl^A_s |\varphi\rangle &=& 0 \,, \qquad  \qquad s=1,2,\ldots
\\
\tjl^3_0 |\varphi\rangle &=& \tm |\varphi\rangle \,.
\ee
Since in the spectral flowed frame $F$, the global $SL(2,R)$ algebra is
\be
j_0^{\pm} &=& \tjl_{\mp w}^{\pm}\, \qquad  j_0^3 = \tjl^3_0 + \frac{k'}{2}w \, \qquad k'=k+2\,,
\ee
the state $|\varphi\rangle$ obeys, for $w$ positive,
\be
j_0^- |\varphi\rangle = 0\, \qquad j_0^3 |\varphi\rangle =
(\tm+\frac{k'}{2}w) |\varphi\rangle \qquad w>0 \,.
\ee
Therefore, in the $F$ frame, $|\varphi \rangle$  is the lowest weight
of a discrete
$d_H^+$ representation of the global algebra, with
spin $H= \tm + \frac{k'}{2}w$. Similarly, for negative $w$, $|\varphi \rangle$
is the highest weight of a discrete representation  $d_H^-$ of the
global algebra, with
spin $H= - \tm - \frac{k'}{2}w$.
We will assume that  for $w>0$ we have $\tm>0$, and for $w<0$ we had $\tm<0$.

\begin{figure}[t]
\begin{center}
  \includegraphics[height=\textwidth,angle=270]{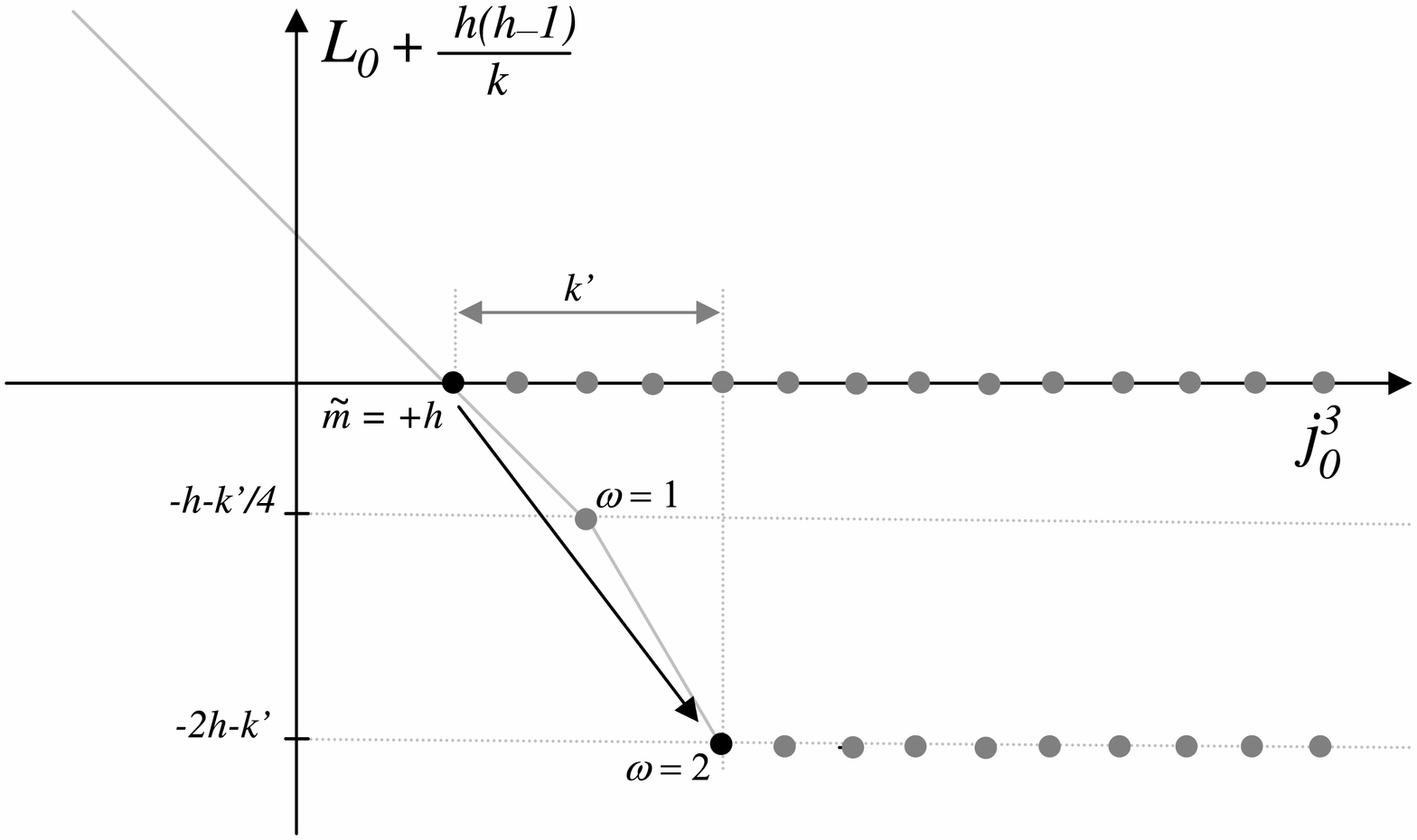}\\
  \caption{Weight diagram of $SL(2,R)_{k'}$. The points in the $j_0^3$
axis are affine primaries in a $d_h^+$ representation.
The spectral flow of the state with  $\tm=h$  by $w=2$ units gives a
state at level~$-2h-k'$,
 which is the lowest weight state of a $d_H^+$ representation of the
global algebra with~$H=h + k'$.
All the states of this $d_H^+$ representation are Virasoro primaries.
  }\label{GraficoSL2}
  \end{center}
\end{figure}

Let us consider the case $w>0$. On the state $|\varphi\rangle$,
we can act with $j_0^+$ to create the infinite higher states of a
discrete lowest weight representation,
which we normalize as
\be
j_0^3 |H,m \rangle &=& m |H,m \rangle \qquad \qquad \qquad  \qquad m =
H, H +1 \ldots
\\
j^{\pm}_0 |H,m \rangle &=& (m \mp (H-1) ) |H,m\pm1 \rangle \, ,
\ee
where $H = \tm+\frac{k'}{2}w$ and $|H,H\rangle = |\varphi \rangle$.
In Figure \ref{GraficoSL2} we show an example of the position of these
new multiplets in the
$SL(2,R)_{k'}$ weight diagram.

We have considered only the holomorphic sector of the theory, but
there is a similar anti-holomorphic
copy of the currents on which an identical amount of spectral flow
must be performed, and the flowed states
depend also on indices $\bH, \bm$.
The operators $\Phi_{m, \bm}^{w}$ that create the flowed modes from
the vacuum can be formally summed into
the field
\be
\Phi_{H, \bH}^w(x, \bx) = \sum_{m,\bm} \Phi_{m; \bm}^{w} \, x^{-H-m}
\bx^{-\bH-\bm} \,,
\label{phiexp}
\ee
with
\be
H &=&  \tm+\frac{k'}{2}w \qquad \qquad
\bH = \tilde{\bar{m}} +\frac{k'}{2}w \,.
\ee
This field is  not an affine primary in the
flowed frame $F$, but it  is still a Virasoro primary.
To see this, note that the positive Virasoro modes in the $F$ frame,
\be
L^H_n &=& \tilde{L}^H_n - w \tjl_n^3 \,,
\ee
annihilate the state $|\varphi \rangle$, and the operator $j^+_0$ which
creates the other modes commutes with~$L_n^H$.
The zero modes of the  $SL(2,R)$ currents act on $\Phi_{H, \bH}^w(x, \bx)$ as
\be
j^A_0 \Phi_{H, \bH}^w(x, \bx) = - D^A_x \Phi_{H, \bH}^w (x,\bx)\,,
\label{zero-modes-on-sl2}
\ee
where $D^A_x$  are the differential operators (\ref{dm})-(\ref{dp})
with $h \rightarrow H$, and
similarly for the anti-holomorphic sector.
Note that we have not mentioned the \sl\ spin $h$ of the unflowed
representation,
since the spin $H$ in the flowed frame $F$ depends only on the value of ~$\tm$.
On the other hand, the conformal dimension of $\Phi_{H, \bH}^w(x, \bx)$ does
depend on $h$ and is given, from~(\ref{lzh}), by
\be
\Delta = -\frac{h(h-1)}{k'-2} -w\tilde{m} - \frac{k'w^2}{4}
\ee
The expression (\ref{phiexp}) for $\Phi_{H, \bH}^w(x, \bx)$ is actually quite
schematic. The field should be considered as a meromorphic function of
$H, \bH$, and its modes
in the  $m,\bm$ basis are obtained from the integral transform
\be
 \Phi_{m, \bm}^{w}= \int \frac{dx^2}{|x|^2} \,\, x^{H+m} \bx^{\bH+\bm}
\Phi_{H, \bH}^w(x, \bx) \,.
\ee
So negative values of $m,\bm$ are obtained also from $\Phi_{H, \bH}^w(x, \bx)$.
Moreover the sign of $w$ is correlated with that of $m, \bm$, so
$\Phi_{H, \bH}^w(x, \bx)$ contains both signs of the spectral flow
parameter $w$,
and we can use positive $w$ to denote
it~\cite{Maldacena:2001km}. A special case   occurs when the original
state was the lowest weight of a discrete
representation $d^+_h $, with $\tm=h$.
In this case  performing spectral flow with $w=-1$ leads to an
{\it unflowed} lowest weight
representation~$d^-_{\frac{k}{2}-h}$~\cite{Maldacena:2000hw}.
Thus, the field in the~$x$ basis contains the representations with
spectral flow parameters~$w$ and~$-w-1$.
This case will be relevant for us in the flowed chiral operators,
and we will denote the operators obtained from $\tilde{m}=
\bar{\tilde{m}}=h$, which
have  $H=\bar{H}= h+k'w/2$  by
\be
\Phi_{h,w}(x,\bar{x})
\label{phiextremal}
\ee
instead of $\Phi_{H, \bH}^w$.

We refer the reader to
\cite{Maldacena:2000hw,Maldacena:2000kv,Maldacena:2001km} for more
details on the $SL(2,R)$ spectral flow.
We now turn to study how the  spectral flow acts in the bosonic
$SU(2)_{k''}$ sector.

\subsection{Spectral Flow in Bosonic $SU(2)$}
Let us see  how the map of an affine primary into a lowest/highest weight of the global algebra
works for the bosonic $SU(2)_{k-2}$ algebra $k^a$.
If $|\varphi\rangle$ satisfies
\be
\tkl^a_s |\varphi\rangle &=& 0 \,, \qquad  \qquad s=1,2,\ldots
\\
\tkl^3_0 |\varphi\rangle &=& \tn |\varphi\rangle \,,
\ee
then, using
\be
k_0^{\pm} &=& \tkl_{\mp w}^{\pm}\,, \qquad  k_0^3 = \tkl^3_0 - \frac{k''}{2}w \,,
\ee
we get that in the flowed frame $F$, for  $w$ positive,
\be
k_0^- |\varphi\rangle = 0\,, \qquad k_0^3 |\varphi\rangle = (\tn-\frac{k''}{2}w) |\varphi\rangle \, \qquad
k''=k-2 \,.
\ee
Thus $|\varphi \rangle$  becomes the lowest weight state of a spin $J=-\tn+\frac{k''}{2}w$ representation of the
global~$SU(2)$. Similarly, for negative $w$, $|\varphi \rangle$ is the highest weight state
of a representation with spin $J= \tn-\frac{k''}{2}w$.

As already mentioned, the spectral flow maps the  Hilbert space of the
$SU(2)$ WZW model to itself. This can be seen from the characters.
In the $\tT$ frame, a spin $j$  character of $SU(2)_{k-2}$ is
\be
\mathrm{Tr}(q^{\tilde{L}_0-\frac{c}{24}} p^{\tkl^3_0} ) = \lambda_l(q,p) =
\frac{\Theta_{l+1,k}(q,p) - \Theta_{-l-1,k}(q,p)}{\Theta_{1,2}(q,p) - \Theta_{-1,2}(q,p)} \, ,
\ee
where $l=2 j=0, \ldots, k\!-\!2$ and
\be
\Theta_{m,k}(q,p) = \sum_{n\in \mathbb{Z}+ \frac{m}{2k}} q^{kn^2} p^{-kn} \,.
\ee
Expressing $L_0,k^3_0$ in terms of $\tilde{L}_0, \tkl_0^3$,
we obtain the corresponding character in the spectral flowed frame $F$ \cite{Feigin:1998sw},
\be
\mathrm{Tr}(q^{L_0-\frac{c}{24}} p^{k^3_0} ) &=& q^{\frac{(k-2)w^2}{4}} p^{-\frac{(k-2)w}{2}} \lambda_l(q,q^{-w}p) \\
&=&
\left\{
\begin{array}{ll}
\lambda_l(q,p)     & \quad \textrm{for} \,\,\,\,\, w \in 2\mathbb{Z}  \\
\lambda_{k-2-l}(q,p) & \quad \textrm{for} \,\,\,\,\, w \in 2\mathbb{Z}+1 \,.
\end{array}
\right.
\label{character-map}
\ee
So a spin $j$ representation is mapped to a spin $j$ or $\frac{k''}{2}-j$ representation, according to
whether~$w$ is even or odd. A similar mapping of the characters for the $SL(2,R)$ algebra can be
found in~\cite{Pakman:2003kh}.

\begin{figure}[t]
\begin{center}
\includegraphics[height=\textwidth,angle=270]{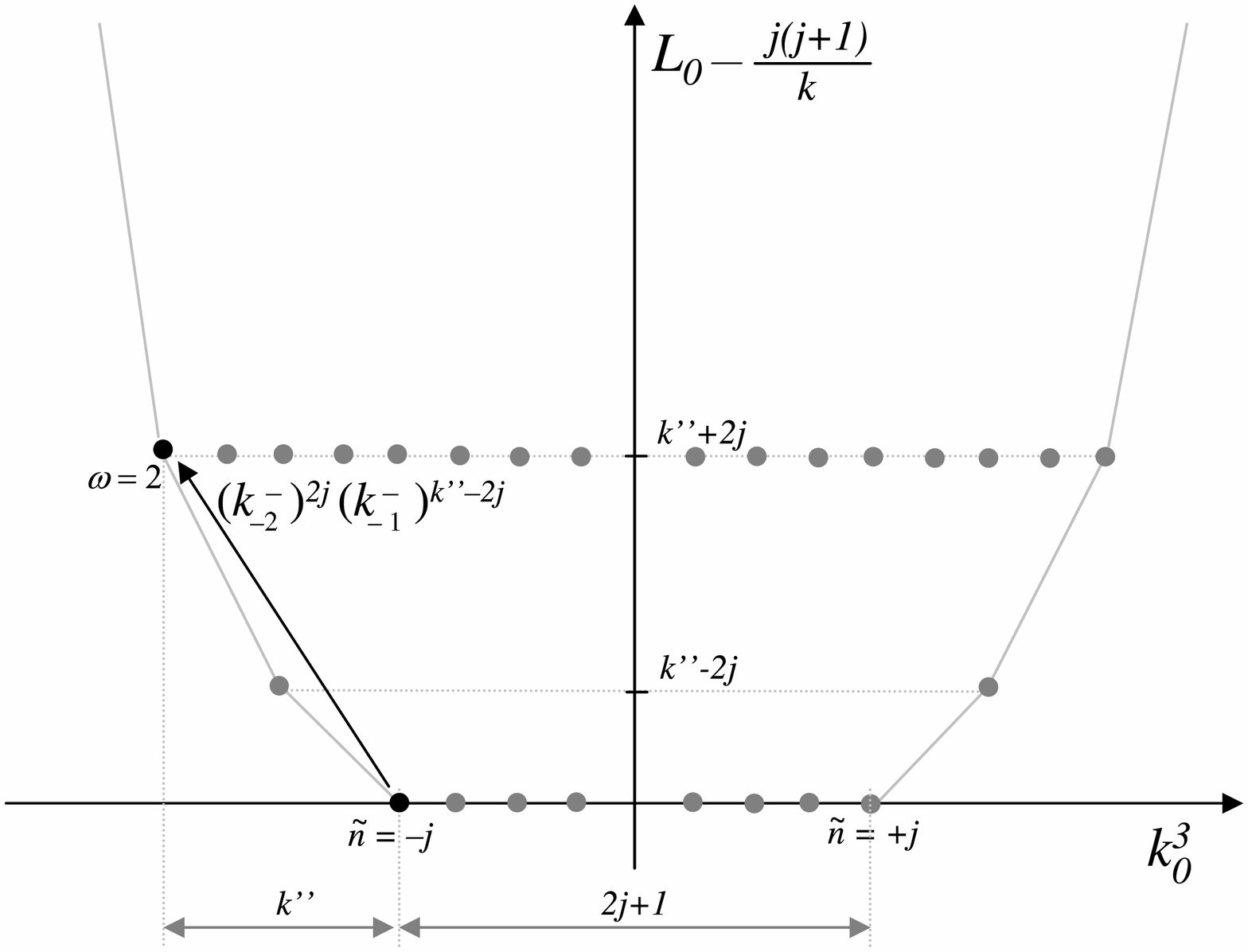}\\
\caption{Weight diagram of a spin $j$ representation of affine $SU(2)_{k''}$ and
the multiplet obtained from spectral flowing the $k_0^3 =-j$ affine primary  by $w=2$ units. The points in the $k_0^3$ axis are~$2j\!+\!1$ affine primaries.
The affine descendants of the horizontal line  at level $k''+2j$ are  Virasoro primaries which form a representation
of the global $SU(2)$ symmetry with spin $J=j+k''$.}
\label{GraficoSU2}
  \end{center}
\end{figure}

In the next section,  we will be interested  in
the case when the original unflowed state is an affine primary $|\!-\!j,j\rangle$,
with spin $j$,  $\tn=-j$ and  $\Delta=j(j+1)/k$.
Let us consider $w$ positive.
For $w$ even, $w=2p$, we claim that this state is mapped into
\be
|\!-\!J, J  \rangle
\equiv (k^-_{-2p})^{2 j}(k^-_{-2p+1})^{k''\!-\!2j} \ldots (k^-_{-2})^{2j}(k^-_{-1})^{k''\!-\!2j}|\!-\!j,j\rangle
\label{flowed-su2-state}
\ee
with
\be
J = j+ k''p \, .
\ee
For $w$ odd, $w=2p+1$,  into
\be
|\!-\!J, J  \rangle
\equiv (k^-_{-2p-1})^{2 j}(k^-_{-2p})^{k''\!-\!2j} \ldots (k^-_{-2})^{k''-2j}(k^-_{-1})^{2j}|\!-k''/2+j, k''/2-j \rangle
\label{flowed-su2-state-odd} \, ,
\ee
with
\be
J = j+ k''(p + 1/2) \,.
\ee
To see that (\ref{flowed-su2-state}) and (\ref{flowed-su2-state-odd}) are the correct states,
it is sufficient to note that their quantum numbers are
\be
k_0^3 &=& -j -wk''/2 \,,
\\
L_0 &=& \frac{j(j+1)}{k} + jw + \frac{k''}{4}w^2 \, ,
\label{flowed-su2-conformaldim}
\ee
as expected from  (\ref{kz}) and (\ref{lzs}) (with $k \rightarrow k''$).
For fixed~$w$ and~$k''$, there is a one-to-one correspondence between
the quantum numbers~$L_0, k_0^3$ of the unflowed and the flowed states, hence
 the multiplicity of the states is preserved under the spectral flow. Since
the original state $|\!-\!j,j\rangle$ was the only one with its quantum numbers,
this guarantees that
~(\ref{flowed-su2-state}) and~(\ref{flowed-su2-state-odd})
are the correct states. To verify that the flowed state is a Virasoro primary,
note that it lies in the border of the weight diagram of the affine $SU(2)_{k''}$ algebra (see Fig. \ref{GraficoSU2}),
so the action of any Virasoro mode~$L_s$ with positive $s$ would take it outside of the diagram.

The full multiplet with spin
\be
J= j + \frac{k''w}{2}
\ee
can be generated by acting on~(\ref{flowed-su2-state}) and (\ref{flowed-su2-state-odd})
with~$k^+_0$. Let us normalize the states in the multiplet  as
\be
k_0^3 |n, J  \rangle &=& n |n, J \rangle \,,
\\
k^{\pm}_0 |n, J \rangle &=& (\pm n +1 + J ) |n \pm 1, J \rangle \,,
\\
k^{\pm}_0 | \! \pm  \!J , J \rangle &=& 0 \,.
\ee
Since the  $k^a_0$'s commute with the Virasoro generators, all the members of the multiplet
are Virasoro primaries with conformal weight (\ref{flowed-su2-conformaldim}).
In Figure \ref{GraficoSU2} we ilustrate the position of the elements of this multiplet in the $SU(2)_{k''}$
weight diagram for the case $w=2$.

Applying the same amount of  spectral flow to the
anti-holomorphic sector, the operators $V^w_{n,\bn}$, which create the states $|n,\bn, J  \rangle$
from the vacuum can be summed into
\be
V_{j,w}(y,\by) &=& \sum_{n,\bn=-J}^{J} V^w_{n,\bn} y^{-n+J} \by^{-\bn +J} \,\,\,.
\label{yexpansion}
\ee
\ni
This field is not an affine primary of the   $k^a$ currents, but the zero modes act on it as
\be
k^a_0 V_{j,w}(y,w) = -  P^a_y V_{j,w}(y,w)
\label{zero-modes-on-su2}
\ee
where $P^a_y$ are the differential operators (\ref{pm})-(\ref{pp}),
with $j \rightarrow J$,
and similarly for the anti-holomorphic currents.

\vskip .5cm

\ni
In the table below, we summarize the quantum numbers of the $SL(2,R)_{k'}$ and $SU(2)_{k''}$
states before and after performing spectral flow by $w$ units, with $w >0$.

\begin{center}
\renewcommand{\baselinestretch}{5.3}
\begin{tabular}{|c||c|c|}
\hline
 & $SL(2,R)_{k'}$ & $SU(2)_{k''}$ \\
\hline \hline
$\tjl_0^3$/$\tilde{k}_0^3$ &  $\tilde{m}$  & $\tilde{n}$    \\
\hline
$j_0^3$/$k_0^3$ &  $\tilde{m} + k'w/2 $  & $\tilde{n} - k''w/2 $    \\
\hline
$H/J$ & $\tilde{m} + k'w/2$ & $-\tilde{n} + k''w/2$ \\
\hline
$\Delta$ & $-h(h-1)/(k'-2) -w\tilde{m} - k'w^2/4 $ &  $j(j+1)/(k''+2) -w\tilde{n} + k''w^2/4$ \\
\hline
\end{tabular}
\linebreak
\vskip 0.4cm
{\bf Table 1}: Quantum numbers in
$SL(2,R)_{k'}$ and $SU(2)_{k''}$  \\ before  and after  performing spectral flow by $w>0$ units.
\end{center}

\section{Spectral Flow for the Free Fermions \label{sfff}}
As already mentioned, the spectral flow for the free fermions is just a rearrangement
of the spectrum. The effect of this rearrangement is to provide finite dimensional
representations of the global $SL(2,R)$ and $SU(2)$ algebras in terms of Virasoro primaries of the~$c=\frac32$ theories of three fermions.
For the~$SU(2)$ case, this construction was  studied in \cite{Aldazabal:1992ae} using an~${\cal N}=1$ theory of a free boson and a free fermion.
This is the supersymmetric version of the construction in~\cite{Witten:1991zd}.

For some expressions, it is convenient to have a bosonized form of the
fermions.  For this, we define
\be
\d H_1 &=& \frac{2}{k} \psi^2 \psi^1 \,, \\
\d H_2 &=& \frac{2}{k} \chi^2 \chi^1 \,, \\
\d H_3 &=& \frac{2}{k} i \psi^3 \chi^3 \,.
\ee
We normalize the  four fermions of $T^4$, $\eta^i, \, i=1\ldots4$, as
\be
\eta^i(z)\eta^j(w) \sim \frac{\delta^{ij}}{z-w}\,,
\ee
and they can be bosonized  as
\be
\d H_4 &=& \eta^2 \eta^1 \,,
\\
\d H_5 &=& \eta^4 \eta^3 \,.
\ee
where
\be
H_{i}(z)H_{j}(w) \sim -  \delta_{ij} \log(z-w) \,.
\ee
In order to get the correct anticommutation
among the fermions in their bosonized form, we should
also introduce  proper cocycles  \cite{Kostelecky:1986xg}.
For that, we first define the number operators
\be
N_{i} = i \oint \d H_{i} \,,
\ee
and then work in terms of bosons redefined as
\be
\hat{H}_i = H_i + \pi \sum_{j<i} N_j \,.
\ee
The fermions are expressed in terms of $\hat{H}_i$ as
\be
e^{\pm i \hat{H}_1} &=& \frac{\psi^1 \pm i \psi^2}{\sqrt{k}} \qquad
e^{\pm i \hat{H}_2} = \frac{\chi^1 \pm i \chi^2}{\sqrt{k}} \qquad
e^{\pm i \hat{H}_3} =  \frac{\chi^3 \mp  \psi^3}{\sqrt{k}} \,\,,
\ee
and the cocycles pick the right signs  using the relation
\be
e^{iaN_{j}} e^{ibH_{j}} =  e^{ibH_{j}}  e^{iaN_{j}} e^{iab} \qquad
\qquad j=1\ldots 3\,\,.
\ee
In terms of the~$\hat{H}_i$ bosons, the fermionic currents are
\be
\hat{\jmath}^3 &=& i \d \hat{H}_1 \,, \\
\hat{\jmath}^{\pm} &=& \pm e^{\pm i \hat{H}_1} \left(  e^{- i \hat{H}_3} - e^{+ i \hat{H}_3}
\right) \,, \\
\hat{k}^3 &=&  i \d \hat{H}_2 \,, \\
\hat{k}^{\pm} &=& \mp e^{\pm i \hat{H}_2} \left(  e^{- i \hat{H}_3} + e^{+ i \hat{H}_3}  \right) \,.
\ee

\subsection{\sl\ Fermionic Multiplets }
\ni
Let us consider the spectral flow in the $\psi^A$ sector first.
The NS vacuum in the~$\tT$ frame is  an excited state in the~$T$
frame, given, for positive $w$, by \cite{Pakman:2003cu}
\be
|\tz \rangle &=&  k^{-w/2} \,\,    \psi_{-w +1/2}^- \psi_{-w +3/2}^- \cdots  \psi_{-1/2}^- \vac
\label{flowed-vac-positive}
\ee
and for negative $w$
\be
|\tz \rangle &=&  k^{-w/2} \,\, \psi_{-|w| +1/2}^+ \psi_{-|w| +3/2}^+ \cdots  \psi_{-1/2}^+ \vac
\ee
where the factor $k^{-w/2}$ is to have $\langle \tz| \tz \rangle = \langle 0| 0 \rangle = 1$.
To check that this representation of $|\tz \rangle$ in the $F$ frame is correct,
note that  it is annihilated by $\tp^{\pm}_n = \psi^{\pm}_{n \pm w}$ and $\tp^3_n=\psi^3_n$ for~$n>0$ and that
it has~$\hat{\jmath}^3_0=-w$ and~$L_0=\frac{w^2}{2}$, as expected from~(\ref{j3h}) and~(\ref{lzh}) with $k=-2$.
As we discussed above, for positive $w$, this is the lowest weight state in a
representation of~$\hat{\jmath}_0^{\pm,3}$ with spin $H = \!- w$,
which in this case is finite dimensional.
So let us call this state
\be
|\tz \rangle = |\!-\!w\rangle\,.
\ee
Starting from it  we can build the whole multiplet,
and we normalize its  $2w+1$ states as
\be
\hat{\jmath}^{3}_0|m \rangle &=& m |m \rangle
\\
\hat{\jmath}^{\pm}_0|m \rangle &=& (m \mp (H-1) )|m \pm 1\rangle
\label{sl-multiplet1}
\\
\hat{\jmath}^{\pm}_0 |\!\pm \! w \rangle &=& 0
\label{sl-multiplet2}
\ee
with $H=-w$.
Let us call $U^w_m$ the fields that create these states from the vacuum.
The lowest and highest states have the simple bosonized expression
\be
U^w_{-w} &=& e^{-iw  \hat{H}_1} \\
U^w_{w} &=& e^{iw  \hat{H}_1}
\ee
and since $\hat{\jmath}^{3}= i \partial \hat{H}_1$, it is easy to see
that they have  the correct quantum numbers.
We can now formally sum   the multiplet into a field $\psi_w(x)$,
\be
\psi_w(x)  = \sum_{m=h}^{-h }x^{-H-m} U_m^w
\label{pswdef}
\ee
with $H=-w$. Note that it has fermion number $(-1)^{w}$.
For example, for  $w=1$ we get
\be
\psi_{w=1}(x) \sim - \psi^+ + 2x\psi^3 - x^2\psi^- \,,
\label{pwone}
\ee
which is the field called $\psi(x)$  in \cite{Dabholkar:2007ey}.
From (\ref{sl-multiplet1}) it follows that the zero modes of the currents
act on $\psi_w(x)$  as
\be
 \hat{\jmath}^A_0 \psi_w(x)  = -D_x^A \psi_w(x)
\ee
where  $D_x^A$ are the differential operators (\ref{dm})-(\ref{dp}) with $H=-w$.

One can   repeat the same exercise  for  the three other affine primaries of the $\tT$ frame in the NS sector,
namely,~$\tp^+_{-1/2}|\tz\rangle, \tp^-_{-1/2}|\tz\rangle$ and $\tp^3_{-1/2}|\tz\rangle$.
We will be interested below in the last two cases.

Consider first the state $\tp^-_{-1/2}|\tz\rangle$ for positive
$w$. From (\ref{flowed-vac-positive}) and $\tp^-_{-1/2} = \psi^-_{-w-1/2}$,
we have
\be
 k^{-1/2}   \tp^-_{-1/2}|\tz\rangle =  k^{-(w+1)/2} \,\,     \psi_{-w -1/2}^- \psi_{-w +1/2}^-  \cdots  \psi_{-1/2}^- \vac
\ee
so this state would have come from the $\tT$ vacuum $|\tz\rangle$ if we had flowed~$w+1$ units instead
of~$w$. It gives rise to an  $SL(2,R)$ multiplet with spin $H=-w-1$,
which, summed as in (\ref{pswdef}), gives the field $\psi_{w+1}(x;z)$.

Consider now, for positive $w$, the state
\be
\sqrt{\frac{2}{k}} \tp^3_{-1/2}|\tz\rangle = \sqrt{2}  k^{-(w+1)/2}  \psi^3_{-1/2} \psi_{-1/2}^- \psi_{-3/2}^- \cdots \psi_{-w +1/2}^-\vac
\ee
In the $F$ frame, it is the lowest state in a representation of the global \sl\ with spin~$H=-w$,
and conformal dimension $\Delta=\frac{w^2}{2}+ \frac12$. We can now build the whole
multiplet as in~(\ref{sl-multiplet1})-(\ref{sl-multiplet2}). Let us call $U^{3,w}_m$ to the operators.
We can and sum over $x$ as in (\ref{pswdef}), and
we call the corresponding field $\psi_w^3(x;z)$. Note that it has fermion number $(-1)^{w+1}$.
For $w=1$ we get
\be
\psi_{w=1}^3(x;z) \sim - \hat{\jmath}^+(z) + 2x \hat{\jmath}^3(z) - x^2\hat{\jmath}^-(z)
\ee
which is the field   called $\hat{\jmath}(x)$ in \cite{Dabholkar:2007ey}.
Finally, consider the state
\be
\frac{\sqrt{2}}{k} \tp^3_{-1/2} \tp^-_{-1/2} |\tz\rangle = \frac{1}{\sqrt{2}}\tilde{\hat{\jmath}}^-_{-1}|\tz\rangle
\ee
Following a reasoning similar to the $\tp^-_{-1/2} |\tz\rangle$ case,
the corresponding field in the flowed frame~$F$ is  $\psi_{w+1}^3(x;z)$.

\subsection{$SU(2)$ Fermionic Multiplets}

The case of the  $SU(2)$ fermions is similar.
The vacuum in the $\tT$ frame is mapped, for positive $w$, to
a state
\be
|\tz \rangle = |\!-\!w\rangle
\ee
with $k_0^3=-w$, which is the lowest weight of a representation of spin $J=w$
for the zero  modes~$k_0^{\pm,3}$ in the flowed frame. We obtain the other states
in the multiplet as
\be
\hat{k}_0^{3}|m\rangle &=& m|m  \rangle \,,
\\
\hat{k}_0^{\pm}|m\rangle &=& (J+1\pm m)|m \pm 1 \rangle \,,
\\
\hat{k}_0^{\pm}|\! \pm\! w \rangle &=& 0 \,,
\ee
with $J=w$, and the operator  that creates the full multiplet is defined as
\be
\chi_w (y) &=& \sum_{n=-w}^{w} y^{-n+w} \, T_{n}^w
\label{cswdef}
\ee
The case $w=1$ is given by
\be
\chi_{w=1}(y) \sim  -\chi^+ + 2y\chi^3 + y^2\chi^- \,,
\label{cwone}
\ee
which is the field called $\chi(y)$ in \cite{Dabholkar:2007ey}.
The action of the zero modes is now
\be
\hat{k}^a_0 \chi_w(y;0)\vac = -P_y^a \chi_w(y;0)\vac
\label{zmonc}
\ee
where  $P_y^a$ are the differential operators (\ref{pm})-(\ref{pp}) with $J=w$.
The field $\chi_w(y,z)$ is a Virasoro primary with dimension $\Delta = \frac{w^2}{2}$.
Another state which will be useful below is the spectral flow of the state $\tc^{-}_{-1/2}|\tz \rangle$.
By an argument similar as above, in the spectral flowed frame $F$, for $w>0$,  it gives rise
to the field~$\chi_{w+1}(y,z)$.

\subsection{The Ramond Sector}
The Ramond sector gives fermionic representations with half integer spin,
and it is convenient to work with the $SL(2,R)$ and $SU(2)$ together.

We define the  Ramond operators in the unflowed frame,
\be
\widetilde S _{[\epsilon_1\, , \epsilon_2 \, , \epsilon_3]} \equiv
e^{i\frac{\epsilon_1}{2} \widetilde H_1+ i\frac{\epsilon_2}{2} \widetilde H_2 + i\frac{\epsilon_3}{2} \widetilde H_3} \, ,
\label{eee}
\ee
and the corresponding states
\be
|  \epsilon_1 \,  \epsilon_2 \, \epsilon_3 \rangle^{\widetilde{} \;} = \widetilde S _{[\epsilon_1\, , \epsilon_2 \, , \epsilon_3]} |\tilde 0 \rangle \,.
\ee
From Table 1, with $k'=-2$ and $k''=2$, we see that the states $| - \,  - \, \pm \rangle^{\widetilde{} }\;$ have $\hat {\jmath}_0^3 = - w -\frac{1}{2}$, $\hat k_0^3 = w + \frac{1}{2}$
and $\Delta = \frac{3}{8} + w^2 + w$. These quantum numbers fix them uniquely
(up to an overall phase) to be
\begin{eqnarray}
| - \,  - \,  \pm \rangle^{\widetilde{} \; } & = & e^{-i (\frac{1}{2} +w) \hat H_1  - i (\frac{1}{2}+w) \hat H_2 \pm
\frac{i}{2}\hat H_3 } | 0 \rangle  \\& = &
 k^{-w} \chi^-_{-w}\cdots \chi_{-1}^- \, \psi_{-w}^-\cdots \psi^-_{-1}  \, e^{-\frac{i}{2} \hat H_1 -\frac{i}{2} \hat H_2 \pm \frac{i}{2} \hat H_3  } | 0 \rangle \,.
\ee

We will also use the notation
\be
| -   \, - \,   \pm \rangle^{\widetilde{} \;}  = | - w/2 - 1/2 \,, -w/2 -1/2 \,   \rangle_\pm^w \, .
\ee
Acting on these states with raising operators $\hat \jmath_0^+ $ and $\hat k_0^+$ we find the states
$| n_1 \, , n_2 \rangle_\pm^w$,
 normalized as
\begin{eqnarray}
\hat \jmath_0^\pm |n_1 \,, n_2 \rangle_\kappa^w  & = & ( n_1\mp (H-1) )  | n_1 \pm 1 \, , n_2 \rangle_\kappa^w \, ,
\quad H= -w - \frac{1}{2} \\
\hat \jmath_0^+ | + (w + \frac{1}{2} ) ,\, n_2 \rangle_\kappa^w  &  = & 0  \\
\hat k_0^\pm |n_1, \, n_2 \rangle_\kappa^w  & = & ( J+1 \pm n_2)  | n_1, \, n_2 \pm 1 \rangle_\kappa^w \, ,
\quad J= w + \frac{1}{2} \\
\hat k_0^\pm | n_1, \, \pm (w + \frac{1}{2} )  \rangle_\kappa^w  & = & 0\,.
\end{eqnarray}
(The above equations hold with the subscript $\kappa$ separately equal to $+$ or $-$). Finally we are in the position to define the states
\be
| S^\pm_w (x, y)\rangle \equiv \sum_{n_1, n_2= -(w+1/2)}^{w+1/2}  x^{w/2+1/2-n_1} y^{w/2+1/2-n_2}\,
|n_1 \, , n_2 \rangle_\pm^w \,.
\ee

\ni
In the table below we summarize the fields that we have defined in the fermionic sectors. They will
enter  the construction of the 1/2 BPS operators.
\vskip .5 cm
\begin{center}
\begin{tabular}{|l|c|c|c|c|c|c|}
\hline
Field                 & $ H $  & $ J $  & $\Delta$              & Unflowed state  & Fermion Number & Sector
\\ \hline \hline
$\psi_w (x) $         &  $-w $ &  -     & $\frac{w^2}{2}$       & $|\tz\rangle$         &  $w$               &  \\
\cline{1-6}
$\psi_{w+1} (x) $         &  $-w -1 $ &  -     & $\frac{(w+1)^2}{2}$      & $ \tp^-_{-1/2} |\tz\rangle$         &  $w+1$    &           \\
\cline{1-6}
$\psi_w^3(x) $        &  $-w $ &  -     & $\frac{w^2}{2} + \frac12$    & $\tp^3_{-1/2} |\tz\rangle$ & $w+1$  & \\
\cline{1-6}
$\psi_{w+1}^3(x) $        &  $-w-1 $ &  -     & $\frac{(w+1)^2}{2} + \frac12$    & $\tp^3_{-1/2} \tp^-_{-1/2} |\tz\rangle$ & $w$   &
\raisebox{4.9ex}[0pt]{NS} \\
\hline
\hline
$\chi_w(y) $          & -      & $w$      & $\frac{w^2}{2}$             & $|\tz\rangle$ &     $w$      &    \\
\cline{1-6}
$\chi_{w+1}(y) $          & -      & $w+1$      & $\frac{(w+1)^2}{2}$          & $\tc^-_{-1/2}|\tz\rangle$ &     $w+1$  &        \\
\cline{1-6}
$\chi_w^3(y) $        & -      & $w$      & $\frac{w^2}{2} + \frac12$   & $\tc^3_{-1/2}|\tz\rangle$   & $w+1$ &\\
\cline{1-6}
$\chi_{w+1}^3(y) $        & -      & $w+1$      & $\frac{(w+1)^2}{2} + \frac12$   & $\tc^3_{-1/2}\tc^-_{-1/2}|\tz\rangle$   & $w$ &
\raisebox{4.9ex}[0pt]{NS} \\
\hline \hline
$S_{w}^\pm(x,y) $        & $- w-\frac{1}{2} $      & $w+\frac{1}{2}$      & $  \frac{3}{8} + w^2 +w $  &  $\tilde{S}^\pm(x,y)|\tz\rangle$   &  & R  \\
\hline
\end{tabular}
\linebreak
\vskip 0.4cm
{\bf Table 2}: Fermionic multiplets  obtained from $w$ units of spectral flow.
\end{center}

\vskip 1cm

\subsection{Interactions of Fermionic Multiplets}
The fermionic multiplets we defined  are not
primaries of the affine algebra. They are however Virasoro primaries,
and the  zero modes of the $\hat{\jmath}^A$
currents act as $D_x^A$ and $P_y^a$. This is sufficient to fix the $x$, $y$ and $z$ dependence of
their two and three-point functions.
Let us consider the NS sector of the \sl\ multiplets for concreteness. The two-point functions are
\be
\langle \psi_w(x_1;z_1) \psi_w(x_2;z_2) \rangle &=&  \frac{(x_{12})^{2w}}{(z_{12})^{w^2}}
\label{2pfp}
\\
\langle \psi_w^3(x_1;z_1) \psi_w^3(x_2;z_2) \rangle &=&   \frac{(x_{12})^{2w}}{(z_{12})^{w^2 +1}}
\label{2pfp3}
\ee
where the coefficient in the rhs is fixed by  taking
~$x_1 \rightarrow \infty$  in $x_1^{-2w}\psi_w(x_1)$ and $x_2=0$, so that
eq.(\ref{2pfp})  becomes
\be
\langle V^w_{-w}(z_1) V^w_{w}(z_2) \rangle =  \langle e^{-iw  \hat{H}_1(z_1)} e^{iw  \hat{H}_1(z_2)}  \rangle &=&  \frac{1}{(z_{12})^{w^2}}
\ee
and similarly for eq.(\ref{2pfp3}).
The three-point functions of three $\psi$ multiplets are
\be
\langle \psi_{w_1}(x_1;z_1) \psi_{w_2}(x_2;z_2) \psi_{w_3}(x_3;z_3) \rangle
&=& f^{(0)}(w_1,w_2,w_3) \, x_{12}^{w_1+ w_2 -w_3} x_{23}^{w_2+ w_3 -w_1} x_{31}^{w_3+ w_1 -w_2}
\nn
\\
&& \qquad \times  z_{12}^{\Delta_1 + \Delta_2-\Delta_3} z_{23}^{\Delta_2 + \Delta_3-\Delta_1} z_{31}^{\Delta_3 + \Delta_1-\Delta_2} \,,
\label{3pfpsi}
\ee
where $\Delta_i= w_i^2/2$. There are actually four possible combinations of $\psi$ and $\psi^3$ fields,
and we denote their three-point functions as follows
\be
\langle \psi_{w_1}(x_1) \psi_{w_2}(x_2) \psi_{w_3}(x_3) \rangle
&=& f^{(0)}(w_1,w_2,w_3) \,,
\label{eqf}
\\
\langle \psi_{w_1}(x_1) \psi_{w_2}(x_2) \psi_{w_3}^3(x_3) \rangle &=& f^{(1)}(w_1,w_2;w_3) \,,
\label{eqfg}
\\
\langle \psi_{w_1}(x_1) \psi_{w_2}^3(x_2) \psi_{w_3}^3(x_3) \rangle &=& f^{(2)}(w_1;w_2,w_3) \,,
\\
\langle \psi_{w_1}^3(x_1) \psi_{w_2}^3(x_2) \psi_{w_3}^3(x_3) \rangle &=& f^{(3)}(w_1,w_2,w_3) \,.
\ee
We have omitted the dependence on the $x_i$ and $z_i$, which is similar in all the cases.
The functions~$f^{(0)}$ and~$f^{(3)}$ are symmetric in the three arguments, and for~$f^{(1)}$
and~$f^{(2)}$ we have indicated the symmetries $f^{(1)}(w_1,w_2;w_3) =  f^{(1)}(w_2,w_1;w_3)$
and $f^{(2)}(w_1;w_2,w_3) =  f^{(2)}(w_1;w_3,w_2)$ by means of the semicolon.

We want to  compute now the structure constants $f^{(i)}$.
As we mentioned above, the $\psi, \psi^3$ multiplets are a generalization to $c=3/2$ of a similar structure
that organizes Virasoro primaries of $c=1$ into $SU(2)$ multiplets~\cite{Witten:1991zd}. For the latter,
the three-point functions were computed in~\cite{Dotsenko:1992mg}, and our results below are a
generalization of those computations.
But instead of computing the four $f^{(i)}$'s, we will see that it is enough to
compute~$f^{(0)}$ and~$f^{(1)}$, and~$f^{(2)}$ and~$f^{(3)}$ are obtained using supersymmetry.

Consider first  $f^{(0)}$.
Each field $\psi_{w}(x)$ is a sum over modes $U^w_m$, given by (\ref{pswdef}). Taking
~$z_1, x_1 \rightarrow \infty$  and $z_2, x_2=0$ gives
\be
\langle U_{-w_1}^{w_1}(\infty) U_{w_2}^{w_2}(0)  U^{w_3}_{w_1-w_2}(1)  \rangle = f^{(0)}(w_1,w_2,w_3) \,.
\label{fus}
\ee
First note that if $w_1 = w_2 + w_3$, the above expression becomes
\be
\langle e^{-iw_1 \hat{H}_1 (\infty)}  e^{iw_2 \hat{H}_1 (0)}  e^{iw_3 \hat{H}_1 (1)}  \rangle = 1
\ee
and similarly for $w_2 = w_3 + w_1$ and $w_3 = w_1 + w_2$. When none of these extremal cases occur,
we can assume that
\be
w_i < w_j + w_k  \qquad \qquad i\neq j\neq k \qquad  i,j,k=1,2,3 \,.
\ee
Then we have
\be
U^{w_3}_{w_1-w_2}(z) &=& \frac{1}{p!} (\hat{\jmath}_0^+)^p U^{w_3}_{-w_3}(z)
\\
&=& \frac{1}{p!}  \left(\frac{2}{\sqrt{k}} \right)^p \!\! \oint \! du_1 \ldots \oint \! du_p \psi^3(u_1) e^{i\hat{H}_1(u_1)}
\! \ldots \! \psi^3(u_p) e^{i\hat{H}_1(u_p)} e^{-iw_3 \hat{H}_1(z)}
\ee
where
\be
p &=& w_1 - w_2 + w_3 \,\,,
\ee
and $p$ should be even so that the total fermion number of the three-point function is even.
With the above expression for $U^{w_3}_{w_1-w_2}$, eq.(\ref{fus}) becomes
\be
f^{(0)}(w_1,w_2,w_3) &=&
\frac{1}{p!}  \left(\frac{2}{\sqrt{k}} \right)^p \!\! \oint \! du_1 \ldots \oint \! du_p  \langle \psi^3(u_1) \! \ldots \! \psi^3(u_p) \rangle
\\
\nn
&& \qquad \qquad \times \prod_{i=1}^p u_i^{w_2} (1-u_i)^{-w_3} \prod_{i<j}(u_i-u_j)
\ee
The contours of the $u_i$'s, which  surround the point $z=1$, can be deformed to include the point~$z=0$, since
the integrand has no singularities at $z=0$. We can then change the exponents infinitesimally into
\be
f^{(0)}(w_1,w_2,w_3) &=&
\frac{1}{p!}  \left(\frac{2}{\sqrt{k}} \right)^p \!\! \oint \! du_1 \ldots \oint \! du_p  \langle \psi^3(u_1) \! \ldots \! \psi^3(u_p) \rangle
\label{fzero}
\\
\nn
&& \qquad \qquad \times \prod_{i=1}^p u_i^{\alpha} (1-u_i)^{\beta} \prod_{i<j}(u_i-u_j)^{2 \rho}
\ee
where
\be
\rho =1/2 \qquad \alpha = w_2 + \ve \qquad \beta = -w_3 - \ve
\ee
This allows us to further change the contours into the $[0,1]$ segment of the real axis,
\be
&& f^{(0)}(w_1,w_2,w_3) = \left( \frac{\sin(\pi \alpha)}{\pi}  \right)^{p} \frac{1}{p!} \left(\frac{2}{\sqrt{k}} \right)^p
\\
\nn
&& \quad \times \int_0^1 \!\! dt_1 \ldots \int_0^1 \!\! dt_p
\langle \psi^3(t_1) \ldots \psi^3(t_p) \rangle \prod_{i<j} (t_i -t_j)^{2\rho} \prod_{i=1}^p t_i^{\alpha} (1-t_i)^{\beta}
\ee
The above integral was computed in \cite{Kitazawa:1987za, AlvarezGaume:1991bj}. Using eqs. (A.7) and (A.11) in \cite{Kitazawa:1987za}
gives\footnote{Note that in \cite{Kitazawa:1987za}, the solution of (A.10) for the special case
$n'=0$ is not obtained by setting $n'=0$ in~(A.11), but by retaining
the last factor in (A.11). }
\be
 f^{(0)}(w_1,w_2,w_3) &=& \left( \frac{\sin(\pi \alpha)}{\pi}  \right)^{p} \left(\sqrt{2} \right)^p  \,
\prod_{i=0}^{p-1} \frac{\Gamma(1 + \alpha + [\frac{i}{2}]) \Gamma(1 + \beta + [\frac{i}{2}])}
{\Gamma(1 + \alpha + \beta + \frac{p}{2} +[\frac{i}{2}])}
\prod_{i=1}^{p} \frac{\Gamma(i - [\frac{i}{2}])}{\sqrt{2}} \,,
\nn
\\
\ee
which can be expanded as
\be
 f^{(0)}(w_1,w_2,w_3)  &=& (\ve)^p
\prod_{i=0}^{p-1} \frac{\Gamma(1 + w_2 + [\frac{i}{2}]) \,  \Gamma(1  -w_3-\ve + [\frac{i}{2}]) \, \Gamma(i + 1- [\frac{i}{2}]) }
{\Gamma(1 + w_2-w_3  + \frac{p}{2} +[\frac{i}{2}])}
\ee
Using now
\be
\Gamma(1 -w_3-\ve + [i/2 ])  \, \Gamma\left(w_3+\ve - [i/2 ]\right) \sim  \frac{(-1)^{w_3-[\frac{i}{2}] }}{\ve}
\ee
we get, as  $\ve \rightarrow 0$,
\be
 f^{(0)}(w_1,w_2,w_3) &=&
\prod_{i=0}^{p-1} \frac{\Gamma(1 + w_2 + [\frac{i}{2}]) \, \Gamma(i + 1- [\frac{i}{2}]) }
{\Gamma(1 + w_2-w_3  + \frac{p}{2} +[\frac{i}{2}])\, \Gamma(w_3- [\frac{i}{2}]) } \,.
\ee
Since $p$ is even, this expression can be rearranged into
\be
 f^{(0)}(w_1,w_2,w_3) &=& \prod_{i=1}^s \frac{\Gamma^2(w_2 + i ) \, \Gamma^2( i) }{ \Gamma^2(\frac{w_1 + w_2 -w_3 }{2} + i ) \, \Gamma^2(w_3 +1-i)}
\label{fnons}
\ee
where
\be
s  &=& \frac{p}{2} = \frac{ w_1 - w_2 + w_3}{2} \,.
\ee
In order to make the symmetry between the $w_i$'s in (\ref{fnons}) manifest, we can use identities like
\be
\prod_{i=1}^s \Gamma(w_2 + i ) = \prod_{i=1+ w_2}^{\frac{w_1+w_2+w_3}{2}} \Gamma( i )
= \prod_{i=1}^{w_2} \frac{1}{\Gamma( i )} \prod_{i=1}^{\frac{w_1+w_2+w_3}{2}} \Gamma( i ) \,,
\ee
and all the  factors in (\ref{fnons}) get  expressed in terms of the function
\be
R(n) \equiv \prod_{i=1}^{\frac{n}{2}} \Gamma^2( i ) \,,
\ee
defined for $n$ even. This gives finally
\be
 f^{(0)}(w_1,w_2,w_3)= R\left(w \right) \prod_{i=1}^{3} \frac{R\left(w - 2w_i \right)}{R(2 w_i)}
\label{fps}
\ee
where
\be
w = w_1 + w_2 + w_3 \,.
\ee
Note that the final expression for $ f^{(0)}$ is symmetric in the $w_i$'s,
although this was not manifest in the intermediate steps of the computation.

The computation of $f^{(1)}$ follows along the same lines, the only difference being that now~$p$ should be
odd in order to have an even total fermion number. The expression for $f^{(1)}$ is given by an integral
like~(\ref{fzero}), but with an additional insertion of $\psi^3(1)$ in the vev of the $\psi^3$ fermions.
This integral can be computed using eqs.(A.16)-(A.17) of \cite{Kitazawa:1987za}, and leads to
\be
f^{(1)}(w_1,w_2;w_3) = \frac{\Gamma\left(\frac{w+1}{2}\right)}{\Gamma(1+w_3) \Gamma\left(\frac{w_1+w_2-w_3+1}{2} \right) }
R\left(w + 1  \right) \prod_{i=1}^{3} \frac{R\left(w - 2w_i +1 \right)}{R(2 w_i)}
\label{fps-one}
\ee

\ni
In order to compute $f^{(2)}$ and $f^{(3)}$, we can use that the $\psi^A$ fermions
have an $N=1$ supersymmetry structure with supercurrent
\be
G = \left( \frac{2}{k}\right)^{3/2} \psi^1 \psi^2 \psi^3 = -\left( \frac{2}{k}\right)^{1/2} \d H_1 \psi^3
\ee
which relates the multiplets $\psi(x)$ and $\psi^3(x)$ as
\be
-iw \psi^3_w(z,x) &=& \oint dz' G(z') \psi_w(z,x)
\\
iw \psi_w(z,x) &=& \oint dz' (z'-z) G(z') \psi^3_w(z,x) \,.
\label{psi-relation2}
\ee
Expressing $\psi_{w_1}(z_1,x_1)$ inside the correlation function (\ref{3pfpsi}) by means of  (\ref{psi-relation2}),
and changing the contour to encircle $\psi_{w_2}(z_2,x_2)$ and $\psi_{w_3}(z_3,x_3)$, one gets
\be
w_1 f^{(0)}(w_1,w_2,w_3) &=& w_2 f^{(2)}(w_3;w_1,w_2) + w_3 f^{(2)}(w_2;w_3,w_1) \,.
\ee
Doing the same operation but starting with $\psi_{w_2}(z_2,x_2)$ and $\psi_{w_3}(z_3,x_3)$ gives similarly
\be
w_2 f^{(0)}(w_1,w_2,w_3) &=& w_3 f^{(2)}(w_1;w_2,w_3) + w_1 f^{(2)}(w_3;w_1,w_2)\,,
\\
w_3 f^{(0)}(w_1,w_2,w_3) &=& w_1 f^{(2)}(w_2;w_3,w_1) + w_3 f^{(2)}(w_1;w_2,w_3)\,,
\ee
and these three equation can be inverted to yield
\be
f^{(2)}(w_1;w_2,w_3) = \left( \frac{w_2^2 + w_3^2 -w_1^2}{2 w_2 w_3} \right) f^{(0)}(w_1,w_2,w_3) \,.
\ee
One can use similar techniques to express $f^{(3)}$ in terms of $f^{(1)}$. The three-point functions
of the $SU(2)$ multiplets $\chi_w(y), \chi_w^3(y)$, are given also by the functions
$f^{(i)}$ up to trivial phases.

\section{ 1/2 BPS Flowed Spectrum \label{chiral-spectrum}}

The operators $\O_h^{(\epsilon)}(x,y)$ in (\ref{obm})-(\ref{oba}) have well defined  spins, $H=J$, under
the {\it total} currents~$J^A, K^A$, and can be expanded  in powers of $x,y$ as

\be
\O_h^{(-)}(x,y)  &= & e^{-\phi} \sum_{m} \, (\psi \Phi)_{h-1,m} \, x^{-h+1-m}  \,\,  \sum_{n=-j}^{j} V_{j,n} \,y^{-n+j}
\label{ominus-exp}
\\
\O_h^{(+)}(x,y)  &= & e^{-\phi} \sum_{m} \, \Phi_{h,m}\, x^{-h-m} \,\,  \sum_{n=-j-1}^{j+1} (\chi V)_{j+1,n} \, y^{-n + j+1}
\\
\O_h^{(a)}(x,y)  &= & e^{-\frac{\phi}{2}} \,  \sum_{m} \sum_{n=-j -1/2}^{j+1/2} (S\Phi V )_{(h-1/2,m+1/2)  \atop (j+1/2,n+1/2) }
x^{-m-h+1/2} y^{-n+j+1/2} e^{\pm i (\hat{H}_4-\hat{H}_5)}
\label{oa-exp}
\ee
where in all the cases the relation $j=h-1$ holds, and the two signs of $e^{\pm i (\hat{H}_4-\hat{H}_5)}$
correspond to $a=1,2$. The modes~$(\psi\Phi)_{h-1,m}, (\chi V)_{j+1,n}$ and~$(S\Phi V )_{(h-1/2,m+1/2)  \atop (j+1/2,n+1/2) }$
are states in  irreducible representations of the tensor product of $\Phi_{h,m}$ and $V_{j,n}$ with the fermions,
with the indicated spins and $J^3_0, K^3_0$ eigenvalues.
Their explicit form is
\be
(\psi\Phi)_{h-1,m} &=&  - \psi^+ \Phi_{h,m-1} +2 \psi^3 \Phi_{h,m}  - \psi^- \Phi_{h,m+1} \,,
\label{ppexpansion}
\\
(\chi V)_{j+1,n} &=& -\chi^+ V_{j,n-1} + 2\chi^3 V_{j,n} + \chi^-V_{j,n+1}  \,,
\ee
and
\be
 (S\Phi V )_{(h-1/2,m+1/2)  \atop (j+1/2,n+1/2) } |0\rangle &=&   |++\rangle_{-} \Phi_{h,m} V_{j,n}
+ \, |+-\rangle_{-} \Phi_{h,m} V_{j,n+1}
\\ && + |-+\rangle_{-} \Phi_{h,m+1} V_{j,n}
+ |--\rangle_{-} \Phi_{h,m+1} V_{j,n+1}
\nn
\ee
where
\be
|\ve_1,\ve_2 \rangle_{-} =  (i)^{\frac{1-\ve_2}{2}} S_{[\ve_1,\ve_2,-\ve_1 \ve_2]} |0 \rangle \,.
\ee
and the spin fields $S_{[\ve_1,\ve_2,\ve_3]}$  are those of~(\ref{eee}).
The first and second signs in $|\pm \pm \rangle_{-}$  refer to the eigenvalues $\hat{\jmath}_0^3=\pm 1/2$ and
$\hat{k}_0^3=\pm 1/2$.

We are interested in chiral states whose $SL(2,R)$ part belongs
to the spectral flowed representations.
It turns out that in order to keep the BRST invariance and the chirality condition~$H=J$,
the easiest way to proceed is to apply the spectral flow to all the $j^A, \psi^A, k^a, \chi^a$ algebras.

\subsection*{1/2 BPS Flowed Spectrum in the NS Sector}
Let us start with an  $\O_h^{(-)}(x,y)$ operator in the unflowed frame $\tT$.
Since the spectral flow is best defined
on states diagonal in  $\tj^3_0$ and $\tk^3_0$,
we pick a generic term in its $x,y$ expansion~(\ref{ominus-exp}).
Omitting the $e^{-\phi}$ factor, we consider then the operator
\be
(\tp \tP )_{h-1,\tilde{m}}\tV_{h-1,\tilde{n}} 
\label{unflowed-chiral}
\ee
which, according to (\ref{ppexpansion}),  creates on the vacuum the state
\be
\left( 2\tp^3_{-1/2} \tP_{h,\tilde{m}} - \tp^+_{-1/2} \tP_{h,\tilde{m} -1} - \tp^-_{-1/2} \tP_{h,\tilde{m} +1}\right)
\tV_{h-1,\tilde{n}}|\tilde{0} \rangle \,.
\label{unflow-chiral-state}
\ee
Note that we denote the spin in the unflowed frame by $h$.
This is a superconformal primary with $\tilde{L}_0=1/2$ in the $\tT$ frame.
We consider it now in the physical  frame $F$, in which we have performed $w$ units
of spectral flow in both \sl\ and~$SU(2)$, with $w$ positive.
The stress tensor and the supercurrent
in~$T$  are  given by  \cite{Pakman:2003cu}
\be
L_s &=& \tilde{L}_s - w \tj_s^3 - w \tk_s^3  \,,
\label{t-modes}
\\
G_r &=& \tilde{G}_r -w \tp_r^3 -w \tc_r^3 \,.
\label{g-modes}
\ee
Note that the terms $\pm \frac{k}{4}w^2$ in $L_0$ have canceled between $SL(2,R)$ and $SU(2)$.
We should require this state to be chiral, have $L_0 =1/2$ and be annihilated by the positive
modes of~$L_s$ and ~$G_r$.
Imposing  $L_0= 1/2$ we get
\be
\tilde{m}=-\tilde{n} \,.
\label{mn}
\ee
The modes $L_{s>0}$  in (\ref{t-modes}) clearly annihilate (\ref{unflow-chiral-state}).
Regarding the supercurrent $G_r$ in (\ref{g-modes}), the modes $\tilde{G}_{r>0}$ and $\tc^3_{r>0}$ annihilate (\ref{unflow-chiral-state}),
but $\tp_{1/2}^3$ does not annihilate the first term in~(\ref{unflow-chiral-state}). Thus we need that term be to absent,
which only happens when $\tilde{m}=h-1$, since the $\tP_{h,\tm}$ operators belong to a discrete highest weight representation
of $SL(2,R)$ in the unflowed frame~$\tT$.
We have found then  that the state
\be
\tp^-_{-1/2} \tP_{h,h} \tV_{h-1,-(h-1)} |\tz\rangle
\label{unflowed-chiral-physical}
\ee
is a superconformal primary with $L_0=1/2$ in the $F$ frame.
According to our discussion in section \ref{spectral},
it is annihilated by $J^-_0,K^-_0$, i.e.,
is the lowest  weight of a representation of the global
algebra $J^A_0,K^a_0$ in the flowed $F$ frame, with spins
\be
H=J= h-1 + \frac{wk}{2}\,,
\label{chir-minus}
\ee
so the chirality condition is automatically satisfied due to  (\ref{mn}).
To obtain the rest of the states in the multiplet, we act on (\ref{unflowed-chiral-physical}) with
\be
J^+_0 = \hat{\jmath}^+_0 + j_0^+
\label{decjz}
\\
K^+_0 = \hat{k}_0^+ + k_0^+
\label{deckz}
\ee
and sum over $x,y$.
In the $x,y$ basis the full multiplet will be
the product of the operators created by the separate action of $\hat{\jmath}^+_0, j_0^+, \hat{k}_0^+$ and~$k_0^+$.
Note that  all the modes in the  multiplet will  be superconformal primaries with $L_0=1/2$, since~$J^+_0$ and $K^+_0$  commute with~$L_s$ and~$G_r$.

We get thus that the type~$(-)$ physical chiral operator, in the spectral flowed sector $w$, in the $-1$ picture, is
\be
\O_{h,w}^{(-)}(x,y) = e^{-\phi} \O_{h,w}(x,y) \psi_{w+1}(x)\chi_w(y) \,,
\label{opmf}
\ee
where $\psi_{w}(x)$ and $\chi_w(y)$ are  defined in (\ref{pswdef}) and (\ref{cswdef}), and
\be
\O_{h,w}(x,y) \equiv \Phi_{h,w}(x)V_{h-1,w}(y) \,,
\ee
with~$\Phi_{h,w}(x)$ and~$V_{h-1,w}(y)$ the holomorphic parts of the operators (\ref{phiextremal})
and~(\ref{yexpansion}). The field $\O_{h,w}(x,y)$ is a kind of  spectral
flowed version   of $\O_{h}(x,y)$. Its  conformal dimension and spins are
\be
\Delta &=& -w^2 -w \,, \\
H &=& h + \frac{wk}{2}   + w\,,
\\
J &=& h-1 + \frac{wk}{2}   - w \,.
\ee
In the physical operator (\ref{opmf}), it appears combined with the field $\psi_{w+1}(x)\chi_w(y)$, whose
quantum numbers are
\be
\Delta &=& w^2 +w + \frac12 \,, \\
H &=& -w -1  \,,
\\
J &=& w \,.
\ee
Summing the quantum numbers of the bosonic and fermionic operators gives  $\Delta=1/2$ and
the chirality relation (\ref{chir-minus}), as expected.

Note that we could also have started by applying, to the original state (\ref{unflow-chiral-state}),
$w$ units of spectral flow in $SL(2,R)$ and $-w$ units of spectral flow in the $SU(2)$ sector.
In the frame $F$, we would get a highest weight state for $SU(2)$, and after summing over the multiplet
created by $J^+_0, K^-_0$ the final operator would coincide with (\ref{opmf}).

For the computation of the three-point functions, we will need the form of $\O_{h,w}^{(-)}(x,y)$ in the zero picture.
For this we can apply the picture rasing operator $e^{\phi}G$ to (\ref{opmf}). But since~$e^{\phi}G$
commutes with $J^+_0, K^+_0$, it is easier to first change the picture
from $-1$ to $0$ in the mode (\ref{unflowed-chiral-physical}) by acting on it with
$G_{-1/2}$, expressed as in (\ref{g-modes}), and only then generate the full multiplet
with $J^+_0, K^+_0$.

We need to use the following commutators
\be
\{ \tilde{G}_{-1/2}, \tp_{-1/2}^- \} &=& \tilde{J}^-_{-1} = \tilde{\jmath}^-_{-1} + \tilde{\hat{\jmath}}^-_{-1} \\
\left[ \tilde{G}_{-1/2}, \tilde{\Phi}_{h,\tm} \right] &=& \frac{2}{k} \tp^A_{-1/2} \tilde{\jmath}_{A,0} \tilde{\Phi}_{h,\tm}
\\
\left[ \tilde{G}_{-1/2}, \tilde{V}_{j,\tn} \right] &=& \frac{2}{k} \tc^a_{-1/2} \tilde{k}_{a,0} \tilde{V}_{j,\tn}
\ee
and when specializing to $\tm=h, \tn=-j$, we also use
\be
\tilde{\hat{\jmath}}^-_{-1} |\tz \rangle &=& \frac{2}{k} \tp^3_{-1/2} \tp^-_{-1/2} |\tz \rangle
\\
\tp^-_{-1/2} \tp^A_{-1/2} \tilde{\jmath}_{A,0} \tilde{\Phi}_{h,h}|\tz \rangle &=& -h \tp^-_{-1/2} \tp^3_{-1/2} \tilde{\Phi}_{h,h}|\tz \rangle
\\
\tc^a_{-1/2} \tilde{k}_{a,0} \tilde{V}_{j,-j}|\tz \rangle &=& -j\tc^3_{-1/2} \tilde{V}_{j,-j}|\tz \rangle  +
\frac12 \tc^-_{-1/2}\tilde{V}_{j,-j+1}|\tz \rangle
\ee
Collecting all the terms,  the picture  zero operator in the flowed $F$ frame, expressed in terms of unflowed
operators is
\be
&& \left( \tilde{\jmath}_{-1}^- + \frac{2}{k}(1-h_w)\tp^3_{-1/2}  \tp^-_{-1/2}
\right) \tP_{h,h} \tV_{h-1,-(h-1)} |\tz\rangle
\label{omzp}
\\
&& +   \frac{2}{k}(1-h_w) \tc^3_{-1/2} \tp^-_{-1/2} \tP_{h,h} \tV_{h-1,-(h-1)} |\tz\rangle
 + \frac{1}{k} \tc^-_{-1/2} \tp^-_{-1/2} \tP_{h,h} \tV_{h-1,-h+2} |\tz\rangle
\nn
\ee
where
\be
h_w = h + \frac{kw}{2}
\ee
Note that in the last term the unflowed $SU(2)$ primary $ \tV_{h-1,-h}$ has $\tn=-h$, and not $\tn=-h-1$ as in the
rest of the terms. We can act on this state with $J_0^+ ,K_0^+$ and sum over all the states. This gives
finally the operator
\be
{\cal Z}_{h,w}^{(-)}(x,y) = {\cal Z}_{h,w}^{(-,1)}(x,y) + {\cal Z}_{h,w}^{(-,2)}(x,y) \,,
\label{omzpmul}
\ee
where
\be
{\cal Z}_{h,w}^{(-,1)}(x,y) &=& \sqrt{1/k} \, j_{-1-w}(x) \O_{h,w}(x,y) \psi_{w}(x)\chi_{w}(y) \\
&& \quad +  \sqrt{2/k} (1-h_w) \O_{h,w}(x,y) \psi^3_{w+1}(x)\chi_{w}(y) \,,
\nn
\\
{\cal Z}_{h,w}^{(-,2)}(x,y) &=& (-1)^{w+1}\sqrt{2/k} (1-h_w) \O_{h,w}(x,y) \psi_{w+1}(x)\chi^3_{w}(y)
\label{zm2}
\\
&& \quad +   (-1)^{w+1} \sqrt{1/k} \Phi_{h,w}(x)V'_{h-1,w}(y) \psi_{w+1}(x)\chi_{w+1}(y) \,,
\nn
\ee
These two terms of ${\cal Z}_{h,w}^{(-)}$  come from the first and second line of~(\ref{omzp}).
In the second term of~(\ref{zm2}), we denoted by~$V'_{h-1,w}(y)$  the $SU(2)$ multiplet of spin $J=h-2 + kw/2$ obtained by
 spectral flowing the operator $\tV_{h-1,-h+2}$ (instead of $\tV_{h-1,-h+1}$).
We also defined
\be
j_{-1-w}(x) = j_{-1-w}^+ - 2x j_{-1-w}^3 + x^2 j_{-\!1\!-w}^-
\ee
which is a combination of modes of $j^A$ with $H=-1$ under the global $SL(2,R)$ algebra.

The reason for splitting ${\cal Z}_{h,w}^{(-)}(x,y)$ into two terms in (\ref{omzpmul})  is that
both  fermion numbers $F_{\psi}$ and $F_{\chi}$ change by one unit from one term to the other.
Since in  a non-zero correlator the fermion number should be even independently in the  $\psi$ and $\chi$ sectors,
whenever ${\cal Z}_{h,w}^{(-,1)}(x,y)$ is non-zero inside a correlator, the contribution of
${\cal Z}_{h,w}^{(-,2)}(x,y)$ will vanish, and viceversa.

Following the same steps for  the $\O_h^{(+)}$ operators, leads to
the flowed operators
\be \O_{h,w}^{(+)}(x,y) = e^{-\phi} \O_{h,w}(x,y)
\psi_{w}(x)\chi_{w+1}(y) \,,
\label{Ozp}
\ee
where now
\be H=J= h_w =h + \frac{kw}{2}
\ee
In  the zero picture this operator becomes
\be {\cal
Z}_{h,w}^{(+)}(x,y) = {\cal Z}_{h,w}^{(+,1)}(x,y) + {\cal
Z}_{h,w}^{(+,2)}(x,y) \,,
\ee
where
\be
{\cal Z}_{h,w}^{(+,1)}(x,y) &=& \sqrt{1/k} \, k_{-1-w}(y) \O_{h,w}(x,y) \psi_{w}(x)\chi_{w}(y) \\
&& \quad +  \sqrt{2/k} (1-h_w) \O_{h,w}(x,y) \psi_{w}(x)
\chi^3_{w+1}(y) \,, \nn
\\
{\cal Z}_{h,w}^{(+,2)}(x,y) &=& (-1)^{w+1}\sqrt{2/k} (1-h_w)
\O_{h,w}(x,y) \psi^3_{w}(x)\chi_{w+1}(y)
\\
&& \quad +   (-1)^{w+1} \sqrt{1/k} \Phi'_{h,w}(x)V_{h-1,w}(y)
\psi_{w+1}(x)\chi_{w+1}(y) \,.
\nn
\ee
Here
\be
k_{-1-w}(y) = k_{-1-w}^+ - 2y k_{-1-w}^3 - y^2 k_{-\!1\!-w}^-
\ee
is a combination of modes with $J=1$ under the global $SU(2)$ algebra,
and $\Phi'_{h,w}(x)$ is the field  obtained
by  spectral flowing  the operator $\tilde{\Phi}_{h,h+1}$ (instead of $\tilde{\Phi}_{h, h}$ ).

\subsection*{1/2 BPS Flowed Spectrum in the R Sector}

To construct the spectral flowed Ramond 1/2 BPS  operators in the -1/2 picture, we
start from the state
\be \label{start1}
\tilde \phi_{h, h} \tilde V^w_{h-1, 1-h}\,  | - \, -\, - \rangle^{\widetilde {} \;}   \,.
\ee
It is easy to check that in the $F$ frame, this is a superconformal primary with $L_0=\frac{3}{8}$.
Moreover, using (\ref{g-modes}), we see that it is annihilated by $G_0$. This ensures that (\ref{start1})
is in the BRST cohomology. Applying the usual procedure of constructing the multiplet
in the $(x,y)$ basis, and adding the dependence on the $T^4$ twisted fields and on the bosonized ghosts,
 we arrive at the 1/2 BPS physical operators
\be
\O_{h.w}^{(a)} (x, y) = e^{-\phi/2} \O_{h,w}(x,y) S^-_w (x,y) \, e^{\pm \frac{i}{2} (\hat H_4 - \hat H_5)}\,.
\ee
To obtain the physical operators in the -3/2 picture, it turns out that we need to start with the state
\be
\tilde \phi_{h, h} \tilde V^w_{h-1, 1-h}\,  | - \, -\, + \rangle^{\widetilde {} \;}   \, .
\ee
This leads to the operators
\be
{\cal Z}_{h.w}^{(a)} (x, y) = -\frac{\sqrt{k}}{(2h-1+kw)} e^{-3\phi/2} \O_{h,w}(x,y) S^+_w (x,y) \, e^{\pm \frac{i}{2} (\hat H_4 - \hat H_5)}\, .
\ee
Let us now check that these are the correct expressions for the physical operators in the -3/2 picture.
 A short computation gives
\be
G_0 \, |  \O_{h,w}(x,y) S^+_w (x,y)  \rangle =  -\frac{(2h -1 + kw)}{\sqrt{k}} \,  |  \O_{h,w}(x,y) S^-_w (x,y)  \rangle \, ,
\ee
so we see that the operation of picture raising brings us from ${\cal Z}_{h.w}^{(a)} (x, y)$ to $\O_{h.w}^{(a)} (x, y)$.
The relative normalization factor $-(2h -1 + kw)/\sqrt{k}$ appearing between the operators in the -1/2 and -3/2 picture will play
an important role in the following.

\begin{center}
\renewcommand{\baselinestretch}{5.3}
\begin{tabular}{|l|c|l|c|c|c|}
\hline
Op.  & Pc & Expansion & $F_{\psi}$ & $F_{\chi}$ & $H\!=\!J$
\\
\hline \hline
 $\O_{h,w}^{(-)}$  & -1 & $ e^{-\phi} \O_{h,w}\psi_{w+1}\chi_w$  &  $w\!+\!1$ & $w$ &
\\
\cline{1-5}
 ${\cal Z}_{h,w}^{(-,1)}$ & 0 & $\sqrt{1/k} j_{-1-w} \O_{h,w} \psi_{w}\chi_{w}+  \sqrt{2/k} (1-h_w) \O_{h,w} \psi^3_{w+1}\chi_{w}$ & $w$ & $w$ &
\\
\cline{1-5}
 ${\cal Z}_{h,w}^{(-,2)}$ & 0
 & $(-1)^{w+1} k^{-\nicefrac12}\psi_{w+1}( \sqrt{2} (1\!-\!h_w) \O_{h,w} \chi^3_{w} +     \Phi_{h,w}V'_{h-1,w} \chi_{w+1})$
 & $w\!+\!1$ & $w\!+\!1$ & \rb{$h_w\!-\!1 $}
\\
\hline \hline
 $\O_{h,w}^{(+)}$  & -1 &  $e^{-\phi} \O_{h,w}(x,y) \psi_{w}(x)\chi_{w+1}(y) $ & $w$ & $w+1$ &
\\
\cline{1-5}
${\cal Z}_{h,w}^{(+,1)}$  & 0 & $\sqrt{1/k} k_{-1-w} \O_{h,w} \psi_{w}\chi_{w}+  \sqrt{2/k} (1-h_w) \O_{h,w} \psi_{w}\chi^3_{w+1}$ &  $w$ &  $w$ &
\\
\cline{1-5}
${\cal Z}_{h,w}^{(+,2)}$  & 0 &
$ k^{-\nicefrac12}\chi_{w+1}( \sqrt{2} (1\!-\!h_w) \O_{h,w} \psi^3_{w} +     \Phi'_{h,w}V_{h-1,w} \psi_{w+1})$
 & $w\!+\!1$ & $w\!+\!1$
 & \rb{$h_w$}
\\
\hline \hline
$\O_{h,w}^{(a)}$  & $-\frac12$ & $  e^{-\phi/2} \O_{h,w}(x,y) S^-_w (x,y) \, e^{\pm \frac{i}{2} (\hat H_4 - \hat H_5)} $ & & &
\\
\cline{1-5}
${\cal Z}_{h,w}^{(a)}$  & $-\frac32$ &   $ -\sqrt{k}(2h_w-1)^{-1} e^{-3\phi/2} \O_{h,w}(x,y) S^+_w (x,y) \, e^{\pm \frac{i}{2} (\hat H_4 - \hat H_5)}$

 &  & & \rbr{$h_w\!-\!\frac12 $}
\\
\hline
\end{tabular}
\linebreak
\vskip 0.4cm
{\bf Table 3}: Chiral operators in the holomorphic sector with $w$ units of spectral flow.
\end{center}

\subsection{The ADE Series}
The holographic duality that we are considering assumes  the A-series for the modular invariant partition function
of the $SU(2)$ WZW model. It is an important open question what the  ADE classification of the $SU(2)$ modular
invariants~\cite{Cappelli:1986hf, Cappelli:1987xt} corresponds to in the boundary theory.
Here we observe that the construction of 1/2 BPS operators can be carried out consistently also in the D and E cases,
since the mapping of $SU(2)$ representations
under spectral flow~(\ref{character-map}) is consistent with the ADE classification.
Indeed, the level $k''$ and the spins $j$ of the
representations that appear in the diagonal terms of the
D and E modular invariants are ($l=2j$)
\be
\begin{array}{lll}
D_{2t+1}  & \qquad k''=4t & \qquad l = 0,2\ldots k''/2
\\
D_{2t+2} & \qquad k''=4t-2 & \qquad l = 0,2\ldots k''/2
\\
E_6 & \qquad k''=10 & \qquad l =0,3,4, 6,7,10
\\
E_7 & \qquad k''=16 & \qquad l = 0,4,6,8,10,12,16
\\
E_8 & \qquad k''=28 & \qquad l = 0,6,10,12,16,18,22,28
\end{array}
\ee
We see that whenever a representation $l$ appears,  the representation $k''-l$ is also present.
Therefore, much like in the A case that we have described in detail,
each 1/2 BPS operator in  the unflowed sector gives rise to infinitely many flowed operators, one for each positive integer~$w$.

\section{Three-point Functions of 1/2 BPS Flowed Operators\label{chiral-3pf} }

Since all the flowed chiral operators involve the field $\O_{h,w} =
\Phi_{h,w}V_{h-1,w}$, we will be interested in the  product of
\be
\langle \Phi_{h_3,w_3}(x_3) \Phi_{h_2,w_2}(x_2) \Phi_{h_1,w_1}(x_1)
\rangle &=& \frac{C_H(w_i,h_i)} {|x_{12}|^{h_{w_1} + h_{w_2} -
h_{w_3} }  |x_{23}|^{h_{w_2} + h_{w_3} - h_{w_1}} |x_{31}|^{h_{w_3}
+ h_{w_1} - h_{w_2}} } \label{chflowed}
\ee
and
\be \langle
V_{j_3,w_3}(y_3) V_{j_2,w_2}(y_2) V_{j_1,w_1}(y_1) \rangle &=&
C_S(w_i,j_i) \times |y_{12}|^{j_{w_1} + j_{w_2} - j_{w_3} }
|y_{23}|^{j_{w_2} + j_{w_3} - j_{w_1}} |y_{31}|^{j_{w_3} + j_{w_1} -
j_{w_2}} \nn \label{csflowed}
\\
\ee
with $ j_i= h_i-1$, and we have defined \be j_{w_i} = j_i +
\frac{k''w_i}{2} \,. \ee The dependence of these correlators on
$x_i$ and $y_i$ is fixed by the action of the zero
modes~(\ref{zero-modes-on-sl2}) and~(\ref{zero-modes-on-su2}). Since
the fields $V_{j,w}(y)$ are descendants of $SU(2)$ primaries, their
three-point function~$C_S(w_i,j_i)$ can be obtained from those of
the primaries  using standard techniques. We will not perform these
computations in this paper, except for the extremal case $j_{w_3} =
j_{w_2} + j_{w_1}$, which is trivial. Still, we can use the $SU(2)$
tensor product  rule for the $SU(2)$ spins $j_{w_i}$, \be j_{w_i}
\leq j_{w_j} + j_{w_k}     \qquad \qquad i\neq j\neq k \qquad
i,j,k=1,2,3 \,, \ee and the relation \be j_{i} \leq j_{j} + j_{k}
\,, \ee which holds between the primaries, to deduce,
for~(\ref{csflowed}),  the selection rule
\be
w_i \leq w_j + w_k \,.
\label{trianw}
\ee
The three-point functions of the \h3\ model in the unflowed sector
were obtained in
\cite{Teschner:1997ft,Teschner:1999ug}.\footnote{See
also~\cite{Becker:1993at, Giribet:1999ft, Ishibashi:2000fn,
Hosomichi:2000bm, Hosomichi:2001fm, Satoh:2001bi}.} General
three-point functions in the flowed sectors in the $x$ basis are not
known yet, but it was argued in~\cite{Maldacena:2001km} that they
satisfy a selection rule less restrictive than (\ref{trianw}), given
by \be w_i \leq w_j + w_k + 1 \,. \label{trianw-h3}
\ee
As far as we know, the only known flowed three-point function
in the $x$ basis was obtained in~\cite{FZZ, Maldacena:2001km} and corresponds to the case $w_1=1,
w_2=w_3=0$. This is allowed by~(\ref{trianw-h3}), but violates the
relation (\ref{trianw}) which the chiral operators must
obey.\footnote{Several aspects of three-point functions in the
spectral flowed sectors of \h3\ were studied in~\cite{FZZ,
Giribet:2000fy, Giribet:2001ft,Hofman:2004ny, Giribet:2004zd,
Giribet:2005ix, Ribault:2005ms, Giribet:2005mc,Minces:2005nb,
Minces:2007td, Iguri:2007af}. In these works, either the $w_1=1,
w_2=w_3=0$ case was studied in the $x$ basis, or general states were
studied in the $m$ basis. In the latter case, the conservation of
$U(1)$ charge imposes always a relation of the form $\tilde{m}_3 +
k''w_3/2 = \tilde{m}_1 + k''w_1/2 + \tilde{m}_2 + k''w_2/2$. This
extremality condition for the flowed spins $\tilde{m}_i+k''w_i/2$ is
never satisfied in the cases needed for the chiral operators. }

\subsection{Fusion Rules}
The fusion rules of the boundary correlators are (\ref{boundary-fus})
\be
&& n_i \leq n_j + n_k -1 \,, \qquad \qquad i\neq j\neq k \qquad  i,j,k=1,2,3 \,,
\label{boundary-fusion}
\ee
For  unflowed representations, these fusion rules coincide in the
bulk with those of the WZW model.
According to the enlarged bulk-to-boundary dictionary, the  lengths~$n_i$ are
\be
n_i = 2j_i + 1 + kw_i
\label{nj}
\ee
and therefore (\ref{boundary-fusion}) is equivalent to the fusion rules of the bosonic $SU(2)_{k-2}$,
\be
&& j_i \leq j_j + j_k \,,
\label{sufr}
\ee
combined with the rule (\ref{trianw})
\be
w_i \leq w_j + w_k \,.
\ee
which we obtained above.

The above results were expressed in the language of $N=4$, but one can verify that
the agreement holds also for the $N=2$ fusion rules, including the operators of type $a$.

\subsection{String Two-point Functions}
In order to compare bulk and boundary three-point functions,
operators at both sides should be normalized in the same way. In the
chiral operators $\O_{h,w}^{(\e, \eb)}$, both the $V_{j,w}$ and the
fermionic factors, as well as the ghosts, have two-point functions
normalized to $1$, and the only subtlety comes from the $\Phi_{h,w}$
operator. The two-point functions in the \h3\ WZW model diverge as
\be \langle \Phi_{h,w}(x_1) \Phi_{h',w}(x_2) \rangle =
|x_{12}|^{-2h_w} B(h,w) \delta(h-h') \,, \label{2pfh3} \ee and this
divergence comes from the infinite volume of the Killing group in
the target space which leaves invariant the positions $x_1$ and
$x_2$ of the two operators. In the string theory two-point
functions, this infinite is multiplied by a zero coming from
dividing by a similar infinite associated to the Killing group of
the worldsheet, thus leading to a finite string theory two-point
function \cite{Kutasov:1999xu}. Remarkably, the finite result of
this cancelation depends on~$h$. Let us call $\Phi_{h,w}(x_i)S_i \,
(i=1,2)$ to the full operator, were $S_i$ stands for the ghosts,
fermions and $SU(2)$ operators. Then the string theory two-point
function is \be \langle \Phi_h(x_1) S_1  \Phi_h(x_2) S_2
\rangle_{String} = (2h-1+ kw)q_h \,  |x_{12}|^{-2h_w} \,,
\label{string2pf} \ee where \be q_h = - \frac{1}{2k \pi^2}B(h) \,.
\ee Here \be B(h) = - \frac{\nu^{-2h+1}}{\pi b^2} \ga(1-b^2(2h-1))
\ee is the coefficient of the \h3\ two-point function (\ref{2pfh3})
for the unflowed primaries and we assume \be \langle  S_1(1)
S_2(0)\rangle = 1 \,. \ee The expression (\ref{string2pf}) requires
some comments. In the case $w=0$, a detailed derivation
of~(\ref{string2pf}) was given in~\cite{Dabholkar:2007ey} following
ideas of~\cite{Maldacena:2001km} (see also~\cite{Aharony:2003vk}).
We see that, up to $h$-independent factors, the constant from the
cancelation of the infinities is~$(2h-1)$.

In the flowed case, one expects changes both in~$B(h)$ and
in~$(2h-1)$. The former should change because the flowed two-point
function in the $x$ basis of the \h3\ WZW model is the two-point
function in the $\tm$ basis of the original operator in the $\tT$
frame. The explicit form can be found in eq.(5.18)
of~\cite{Maldacena:2001km}, but when $\tm=\bar{\tm}=h$, the
contributions depending on $\tm, \bar{\tm}$ cancel and we get
$B(h,w)=B(h)$. As for the $(2h-1)$ factor, it is shown
in~\cite{Maldacena:2001km} that it changes to $(2h-1+ kw)$ by
introducing a suitable regularization of the divergences. We refer
the reader to Sec 5.1 of~\cite{Maldacena:2001km} for more details.

Using the above result for the string theory two-point functions,
the normalized chiral operators are, in the NSNS sector, \be
\mathbb{O}_{h,w}^{(\e, \eb)}(x,y) &=& \frac{c\, \O_{h,w}^{(\e,
\eb)}(x,y) }{\sqrt{q_h (2h-1+kw)}} \,,
\\
\mathbb{Z}_{h,w}^{(\e, \eb)}(x,y) &=& \frac{c\, {\cal Z}_{h,w}^{(\e,
\eb)}(x,y) }{\sqrt{q_h (2h-1+kw)}} \,, \ee for $\e, \eb= \pm$. The R
sector has an important subtlety. The computation of
(\ref{string2pf}) in the sphere requires the total picture number to
be $-2$. So we can take one of the operators in the $-1/2$ picture
and the other in the $-3/2$. Taking into account that operators in
these two pictures differ by a factor of~$(2h-1+kw)^2/k$, the
string two-point function~(\ref{string2pf}) in the RR sector becomes
\be
\langle \O_{h,w}^{(1,1 )} \O_{h,w}^{(2,2)}  \rangle_{String} =
\frac{k \, q_h}{(2h-1+ kw)} \, |x_{12}|^{-2h_w}
\ee
and therefore the
normalized RR operators  are
\be
\mathbb{O}_{h,w}^{(a, \bar{a})}(x,y) &=& \sqrt{\frac{(2h-1+kw)}{k q_h}}\,c\, \O_{h,w}^{(a,
\bar{a})}(x,y)\,.
\ee
The normalized operators in the R-NS cases are
similarly obtained. Note that we have included also the $c$ ghost as
part of the normalized operators.

\subsection{String Three-point Functions}

\subsubsection{R-R-NS correlators}
The two possible correlators of this type are \be && \langle
\mathbb{O}^{(2,2)}_{h_3,w_3}(x_3,y_3)
\mathbb{O}^{(1,1)}_{h_2,w_2}(x_2,y_2)
\mathbb{O}^{(-,-)}_{h_1,w_1}(x_1,y_1) \rangle
\\
&& =   \left( \frac{(2h_{w_3}-1)(2h_{w_2}-1)}{(2h_{w_1}-1) q_{h_1}
q_{h_2} q_{h_3} }\right)^{\nicefrac12} \frac{g_6}{k} \langle
\O^{(2,2)}_{h_3,w_3}(x_3,y_3) \O^{(1,1)}_{h_2,w_2}(x_2,y_2)
\O^{(-,-)}_{h_1,w_1}(x_1,y_1) \rangle
\\
&& =  \left( \frac{(2h_{w_3}-1)(2h_{w_2}-1)}{(2h_{w_1}-1)
}\right)^{\nicefrac12}
 \frac{ g_6 \, C_H(w_i,h_i) C_S(w_i,j_i) C^2_f(w_i)}{k \sqrt{q_{h_1} q_{h_2} q_{h_3}}}
\label{rrns1} \ee and \be && \langle
\mathbb{O}^{(+,+)}_{h_3,w_3}(x_3,y_3)
\mathbb{O}^{(1,1)}_{h_2,w_2}(x_2,y_2)
\mathbb{O}^{(2,2)}_{h_1,w_1}(x_1,y_1) \rangle
\\
&& =  \left( \frac{(2h_{w_2}-1)(2h_{w_1}-1)}{(2h_{w_3}-1) q_{h_1}
q_{h_2} q_{h_3} }\right)^{\nicefrac12} \frac{g_6}{k}  \langle
\O^{(+,+)}_{h_3,w_3}(x_3,y_3) \O^{(1,1)}_{h_2,w_2}(x_2,y_2)
\O^{(2,2)}_{h_1,w_1}(x_1,y_1) \rangle
\\
&& = \left( \frac{(2h_{w_2}-1)(2h_{w_1}-1)}{(2h_{w_3}-1)
}\right)^{\nicefrac12} \frac{g_6 \, C_H(w_i,h_i) C_S(w_i,j_i)
D^2_f(w_i)}{k \sqrt{q_{h_1} q_{h_2} q_{h_3}}} \label{rrns2}
\ee
where
$j_i= h_1-1$ and we have omitted the  the dependence on $x_i, y_i$,
which is standard. The three-point functions $C_H(w_i,h_i)$ and
$C_S(w_i,j_i)$ were defined in (\ref{chflowed}) and
(\ref{csflowed}), and $C_f$ and $D_f$ are \be
C_f(w_i) &=& \langle S^-_{w_3}(x_3,y_3) \, S^-_{w_2}(x_2,y_2) \, \psi_{w_1+1}(x_1) \chi_{w_1}(y_1)  \rangle \,, \\
D_f(w_i) &=& \langle \psi_{w_3}(x_3) \chi_{w_3+1}(y_3) \,
S^-_{w_2}(x_2,y_2) \, S^-_{w_1}(x_1,y_1)   \rangle \,. \ee In
(\ref{rrns1}) and (\ref{rrns2}), these fermionic couplings
$C_f(w_i)$ and $D_f(w_i)$ appear squared because we include the
holomorphic and antiholomorphic contributions.

For these correlators we will specialize to the $N=2$ extremal
three-point functions, which are the cases computed in the boundary
theory. The extremality  relation is \be J_3 = J_1 + J_2,
\label{jextremal} \ee and the  correlators (\ref{rrns1}) and
(\ref{rrns2}) correspond to the $N=2$ cases \be
\begin{array}{rclclcr}
(\,a\,) & \times & (-) &\rightarrow& (\,a\,) & &      \\
(\,a\,) & \times & (\,a\,) &\rightarrow& (+) & &
\end{array}
\ee respectively. In the first case (\ref{rrns1}), the total spin
for each operator is \be
J_1 &=& j_1 + kw_{1}/2 \\
J_2 &=& j_2 + kw_2/2 + 1/2 \\
J_3 &=& j_3 + kw_3/2 + 1/2 \ee and for the second case (\ref{rrns2})
\be
J_1 &=& j_1 + kw_{1}/2  + 1/2\\
J_2 &=& j_2 + kw_2/2 + 1/2 \\
J_3 &=& j_3 + kw_3/2 + 1 \ee In both cases (\ref{jextremal}) gives
\be w_3 &=& w_1 + w_2 \,, \label{extremw}
\\
j_3 &=& j_1 + j_2 \,. \label{extremj} \ee and combining these
relations with the bulk-to-boundary dictionary \be n = 2j+1 + kw \ee
we get \be n_3=n_1+n_2-1 \,, \ee as in the boundary. In order to get
a precise agreement  between the bulk structure constants
(\ref{rrns1}) and (\ref{rrns2}) and the boundary expressions
(\ref{boundary3pf4}) and (\ref{boundary3pf5}) the following
identities should hold\footnote{Note that even before specializing
to the extremal cases, (\ref{rrns1}) and (\ref{rrns2}) coincide with
the boundary couplings (\ref{boundary3pf4}) and (\ref{boundary3pf5})
if we assume (\ref{pred}). This fact, along with the predictions for
these type of non-extremal correlators  presented
in~\cite{Pakman:2007hn}, suggest that~(\ref{pred}) might hold even
without assuming (\ref{extremw}) and (\ref{extremj}), but we will
only consider the extremal case in this work. }
\be
 \left(\frac{1}{N} \right)^{1/2} = \frac{ g_6 \,  C_H(w_i,h_i) C_S(w_i,j_i) C^2_f(w_i)}{k \sqrt{q_{h_1} q_{h_2} q_{h_3}}} =
\frac{g_6 \,  C_H(w_i,h_i) C_S(w_i,j_i) D^2_f(w_i)}{k \sqrt{q_{h_1}
q_{h_2} q_{h_3}}} \label{pred}
\ee We will now turn these
expressions into a prediction for $C_H(w_i,h_i)$, since the other
factors can be easily computed. Let us start with the fermionic
couplings. The fermionic operators have the expansions
\be \psi_w
(x) &=& \sum_{m=-w}^{w} x^{-m+w} \, U_{m}^w
\\
\chi_w (y) &=& \sum_{n=-w}^{w} y^{-n+w} \, T_{n}^w
\\
S_w^-(x,y) &=& \sum_{m,n=-w-\frac12}^{w+\frac12} x^{-m+w+\frac12}
y^{-n+w+\frac12} \, S_{m,n}^{(-,w)}
\ee
In terms of these modes it
is easy to see that \be C_f(w_i) &=& \langle
S_{-w_3-1/2,-w_3-1/2}^{(-,w_3)}  \,\, \,
S_{w_2-1/2,w_2+1/2}^{(-,w_2)}  \,\,\,  U^{w_1+1}_{w_1+1}
T^{w_1}_{w_1} \rangle \label{cf}
\\
D_f(w_i) &=& \langle  U^{w_3}_{-w_3} T^{w_3+1}_{-w_3-1} \,\,\,
S_{w_2-1/2,w_2+1/2}^{(-,w_2)}  \,\, \,
S_{w_1+1/2,w_1+1/2}^{(-,w_1)}    \rangle \label{df} \ee All the
modes are  either lowest or highest elements of the multiplet,
except for \be S_{w_2-1/2,w_2+1/2}^{(-,w_2)}(0) &=& \hat{\jmath}_0^-
S_{w_2+1/2,w_2+1/2}^{(-,w_2)}(0)
\\
&=& \oint_0 dz e^{- i \hat{H}_1(z)} \left(  e^{+ i \hat{H}_3(z)} -
e^{- i \hat{H}_3(z)}  \right) e^{i(w_2 + \frac12 )(H_1(0) + H_2(0) -
\frac{i}{2}H_3(0)} \,. \nn \ee Inserting this expression into
(\ref{cf}) and (\ref{df}), we get \be
C_f(w_i) = D_f(w_i) &=& \oint_0 dz (z-1)^{w_3} z^{-w_2-1} \\
&=& \frac{w_1! w_2!}{(w_1+w_2)!}
\ee where we used Cauchy's theorem in the last line. The result
$C_f(w_i) = D_f(w_i)$ is a consistency check on the
prediction~(\ref{pred}).

Let us consider now $C_S(w_i,j_i)$. The $k^3$ current can be
bosonized as \be k^3 &=& i\sqrt{\frac{k''}{2}}\d Y \,, \ee with \be
Y(z)Y(w) \sim -\log(z-w) \,, \ee and this allows to represent the
affine unflowed primaries of $SU(2)_{k''}$ as \be V_{j, \tn} =
e^{i\tilde{n} \sqrt{\frac{2}{k''}} Y }\Sigma_{j,\tilde{n}} \,, \ee
where $\Sigma_{j,\tilde{n}}$ are fields in the parafermionic
$SU(2)/U(1)$ theory. In this representation, after spectral flow
with $w>0$ from the $\tn=-j$ state, the lowest/highest weight states
of the  global $SU(2)$ multiplet with spin $j_w = j+wk''/2$ are \be
V_{\pm j_w}^{w} = \left\{
\begin{array}{ll}
e^{\pm ij_w \sqrt{\frac{2}{k''}} Y }\Sigma_{j,\pm j} & \quad \textrm{for} \,\, w  \,\,\textrm{even} \\
e^{\pm ij_w \sqrt{\frac{2}{k''}} Y }\Sigma_{k''/2-j ,\, \pm
(k''/2-j)}  & \quad \textrm{for} \,\, w  \,\, \textrm{odd}
\end{array}
\right. \ee These are the vertex operators that create the states
(\ref{flowed-su2-state}) and (\ref{flowed-su2-state-odd}), and their
highest weight counterparts. From the extremality condition
$w_3=w_2+w_1$, it follows that either all the  $w_i$'s are even, or
two of them are odd. Without loss of generality, we will assume that
in the latter case $w_1$ and $w_2$ are odd. Using the above
representation for $V_{\pm j_w}^{w}$, we get \be C_S(w_i,j_i) &=&
\langle V_{-j_{w_3} }^{w_3} V_{j_{w_2} }^{w_2} V_{j_{w_1} }^{w_1}
\rangle
\\
&=& \left\{
\begin{array}{ll}
\langle \Sigma_{-j_3,j_3} \Sigma_{j_2,j_2} \Sigma_{j_1,j_1} \rangle  & \quad \quad \textrm{for all} \,\, w_i's  \,\,\textrm{even} \\
\langle \Sigma_{-j_3,j_3} \Sigma_{k''/2-j_2,k''/2-j_2}
\Sigma_{k''/2-j_1,k''/2-j_1} \rangle   & \quad \quad \textrm{for}
\,\, w_1, w_2  \,\, \textrm{odd}
\end{array}
\right.
\\
&=& \left\{
\begin{array}{ll}
\langle V_{j_3, -j_3 } V_{j_2, j_2 } V_{j_1, j_1 } \rangle  & \quad  \qquad \textrm{for all} \,\, w_i's  \,\,\textrm{even} \\
\langle V_{j_3, -j_3 } V_{k''/2 -j_2, k''/2 -j_2 } V_{k''/2- j_1,
k''/2- j_1 } \rangle  &  \quad  \qquad  \textrm{for} \,\, w_1, w_2
\,\, \textrm{odd}
\end{array}
\right. \label{csw-cs}
\ee
These identities follow from the fact
that the boson $Y$ is a free field and its contribution to the
correlation functions is trivial. In the extremal case $j_3= j_2 +
j_1$, we have
\be
\langle V_{j_3, -j_3 } V_{j_2, j_2 } V_{j_1, j_1 }
\rangle = C_S(j_3, j_2, j_1)\,.
\ee
This is the three-point function
of the $SU(2)_{k''}$ affine primaries in the $y, \bar{y}$ basis,
given by~\cite{Zamolodchikov:1986bd}\footnote{See
also~\cite{Christe:1986cy, Dotsenko:1990zb}.} \be C_S(j_1,j_2,j_3)=
\sqrt{\ga(b^2)} P(j+1) \prod_{i=1}^{3} \frac{P(j-2j_i)}{ P(2j_i)
\sqrt{\ga((2j_i+1)b^2)} }\,, \label{zf3pf} \ee where \be
j &=& j_1+j_2+j_3 \,, \\
b &= & 1/\sqrt{k} \,, \\
\ga(x) &=& \frac{\Gamma(x)}{\Gamma(1-x)}\,. \ee The function  $P(s)$
is defined for $s$ a non-negative integer as \be P(s) =
\prod_{n=1}^{s} \ga(n b^2)\,, \qquad \qquad P(0)=1\,. \label{ps} \ee
The expression (\ref{zf3pf}) has the remarkable symmetry \be
C_S(j_1,j_2,j_3) = C_S(j_1,k''/2- j_2,k''/2-j_3) \ee and similarly
for any pair of $j_i$'s, as can be seen from the identity \be P(s) =
P(k-s-1) \,. \ee Therefore eq.(\ref{csw-cs}) becomes \be
C_S(w_i,j_i) = C_S(j_i) \,, \ee for any value of the $w_i$'s.

We now have all the elements to go back to~(\ref{pred}) and predict
the three-point function
\be C_H(w_i,h_i) = \frac{\sqrt{q_{h_1}
q_{h_2} q_{h_3}} }{C_S(h_i-1)} \left(
\frac{(w_1+w_2)!}{w_1! w_2!}\right)^{2}
\label{pred2}
\ee
for $h_3=h_1+h_2-1$ and
$w_3=w_2+w_1$. Note that there was a cancelation between $1/\sqrt{N}$ and $g_6/k$,
which in particular makes (\ref{pred2}) independent of $Q_1$, as it should since this is
a statement on the worldsheet CFT, which does not depend on $Q_1$.
The function $C_S(j_i)$ is defined in~(\ref{zf3pf}) for semi-integer
values of the $j_i$'s, but $C_H(w_i,h_i)$ should be well defined for
any values of the $h_i$'s. So we expect that the above equation will
hold when $C_S(j_i)$ is replaced by its extension to continuous
$j_i$'s obtained in~\cite{Dabholkar:2007ey}, given by \be
c_S(a_1,a_2,a_3) =  \frac{\sqrt{\ga(b^2)} b^{\frac12 -b^2}}{\up(b)}
\up (a + 2b) \prod_{i=1}^{3} \frac{\up (a-2a_i +b ) } {[\up (2a_i
+b) \up (2a_i +2b)]^{1/2}} \,, \label{su3pf} \ee where $a_i=bj_i$
and $a = a_1 + a_2 + a_3$. The function $\up$, introduced
in~\cite{Zamolodchikov:1995aa}, is related to the Barnes double
gamma function and can be defined by \be \log \up(x) =
\int_0^{\infty} \frac{dt}{t} \left[\left( \frac{Q}{2} -x \right)^2
e^{-t} - \frac{\sinh^2((\frac{Q}{2}-x)\frac{t}{2})}{\sinh
\frac{bt}{2} \sinh \frac{t}{2b}} \right]\,. \ee The integral
converges in the strip $0<\textrm{Re}(x)< Q$. Outside this range it
is defined by the relations \be \up(x + b) = b^{1-2bx}\ga(bx) \up(x)
\qquad \up(x + 1/b) = b^{-1+2x/b}\ga(x/b) \up(x) \,. \label{shiftu}
\ee Using these  properties, one can verify that $c_S(a_i)$ reduces
to $C_S(j_i)$ for semi-integer~$j_i$'s.

\subsubsection{NS-NS-NS correlators}
The type of predictions that we can make in this case are somewhat
weaker than in the previous section. For example,  let us  consider
three operators of type $(--)$, such that \be w_1 + w_2 + w_3 = \,\,
\textrm{even} \ee In order to have total picture $-2$, we consider a
string three-point function with  two $\mathbb{O}^{(--)}_{h,w}$'s
and one $\mathbb{Z}^{(--)}_{h,w}$. Due to the latter, we will need
the correlator \be \langle \Phi_{h_1,w_1}(x_1)  \Phi_{h_2,w_2}(x_2)
j_{-1-w_3}(x_3)\Phi_{h_3,w_3}(x_3) \rangle &=&
G(h_i,w_i)C_H(w_i,h_i) \,, \ee where the function  $G(h_i,w_i)$
carries the effect of the current algebra descendants, and we omitted the $x_i$'s and
$z_i$'s.
Unfortunately, since the fields $\Phi_{h_i,w_i}(x_i)$  are not
affine primaries, the standard techniques to obtain~$G(h_i,w_i)$
cannot be applied. The three-point function we are interested is
given then by \be && \langle \mathbb{O}^{(--)}_{h_1,w_1}
\mathbb{O}^{(--)}_{h_2,w_2} \mathbb{Z}^{(-,1,-,1)}_{h_3,w_3} \rangle
= \frac{g_6}{k} \prod_{i=1}^3 \frac{1}{\sqrt{q_{h_i}(2h_{w_i}-1)}
}
\\
&& \times  \,\, C_H(h_i,w_i)C_S(h_i-1,w_i) (f^{(0)}(w_1,w_2,w_3))^2
\nn
\\
&& \times \,\, \left( f^{(0)}(w_1+1, w_2+1, w_3)G(h_i,w_i) +
\frac{(2-2h_3)}{\sqrt{2}}f^{(1)}(w_1+1, w_2+2;w_3+1) \right)^2 \nn
\ee where the squares come from  the holomorphic and the
antiholomorphic contributions, and we  omitted the standard
dependence on the $x_i, y_i$. The functions $f^{(0)}$ and $f^{(1)}$
come from the fermion interactions and are given by~(\ref{fps})
and~(\ref{fps-one}).
This expression should coincide with
the first line of the boundary  correlator (\ref{sym}) with $\e_i,
\eb_i=-$, and this implies \be &&   \,\, C_H(h_i,w_i)C_S(h_i-1,w_i)
(f^{(0)}(w_1,w_2,w_3))^2 \nn
\\
&& \times \,\, \left( f^{(0)}(w_1+1, w_2+1, w_3)G(h_i,w_i) +
\frac{(2-2h_3)}{\sqrt{2}}f^{(1)}(w_1+1, w_2+2;w_3+1) \right)^2 \nn
\\
&& = (h_{w_1} + h_{w_2} + h_{w_3} -2)^2 \ee
Similar expressions can
be obtained by considering the other cases.

\section{Conclusions}

We have completed the bulk-to-boundary dictionary for 1/2 BPS
operators in  $AdS_3/CFT_2$, giving concrete expressions for the
physical bulk vertex operators in the flowed sectors, and we have
obtained some partial results about their three-point functions. The
structure of the  string three-point functions (especially for
R-R-NS correlators, where we were able to be more explicit) suggests
that the agreement with the boundary results in \symT\
  holds in the flowed sectors as well. A definite confirmation
  of this expectation must await the evaluation of some missing three-point couplings
  in the $H_3^+$ WZW model, which is an interesting CFT question
  in its own right.  It would be very interesting to see if the
  techniques of \cite{Teschner:1997ft} are effective in this context.

An alternative approach to the
 evaluation of correlation functions in $AdS_3 \times S^3 \times M^4$
may be to exploit the ground ring structure discovered in \cite{Rastelli:2005ph}.
This approach is very efficient in the minimal string~\cite{Kostov:2005av} and
 it would be interesting to see if it can be adapted to this critical background.

\section*{Acknowledgements}
We thank Atish Dabholkar, Juan Maldacena,  Carmen Nu\~{n}ez and Massimo Porrati
for conversations and correspondence.
The work of G.G. is supported by Fulbright
Commission and by Conicet. The work of A.P. is supported by
the Simons Foundation. The work of L.R. is supported in part by the National Science Foundation Grant No. PHY-
0354776 and by the DOE Outstanding Junior Investigator Award.   Any opinions, findings, and conclusions or recommendations
expressed in this material are those of the authors and do not necessarily reflect the views of
the National Science Foundation.

\label{global}

\newpage
\appendix

\newpage
\bibliography{h3bib}

\providecommand{\href}[2]{#2}\begingroup\raggedright\begin{thebibliography}{10}

\bibitem{Maldacena:1997re}
J.~M. Maldacena, {\it The large n limit of superconformal field theories and
  supergravity},  {\em Adv. Theor. Math. Phys.} {\bf 2} (1998) 231--252,
  [\href{http://xxx.lanl.gov/abs/hep-th/9711200}{{\tt hep-th/9711200}}].

\bibitem{Aharony:1999ti}
O.~Aharony, S.~S. Gubser, J.~M. Maldacena, H.~Ooguri, and Y.~Oz, {\it Large n
  field theories, string theory and gravity},  {\em Phys. Rept.} {\bf 323}
  (2000) 183--386, [\href{http://xxx.lanl.gov/abs/hep-th/9905111}{{\tt
  hep-th/9905111}}].

\bibitem{Dijkgraaf:2000vr}
R.~Dijkgraaf, {\it On the d1-d5 conformal field theory},  {\em Class. Quant.
  Grav.} {\bf 17} (2000) 1035--1048.

\bibitem{David:2002wn}
J.~R. David, G.~Mandal, and S.~R. Wadia, {\it Microscopic formulation of black
  holes in string theory},  {\em Phys. Rept.} {\bf 369} (2002) 549--686,
  [\href{http://xxx.lanl.gov/abs/hep-th/0203048}{{\tt hep-th/0203048}}].

\bibitem{Martinec}
E.~Martinec, {\it The d1-d5 system},
  \href{http://xxx.lanl.gov/abs/http://hamilton.uchicago.edu/$\sim$ejm/japan99%
.ps}{{\tt http://hamilton.uchicago.edu/$\sim$ejm/japan99.ps}}.

\bibitem{Gaberdiel:2007vu}
M.~R. Gaberdiel and I.~Kirsch, {\it Worldsheet correlators in ads(3)/cft(2)},
  {\em JHEP} {\bf 04} (2007) 050,
  [\href{http://xxx.lanl.gov/abs/hep-th/0703001}{{\tt hep-th/0703001}}].

\bibitem{Dabholkar:2007ey}
A.~Dabholkar and A.~Pakman, {\it Exact chiral ring of ads(3)/cft(2)},
  \href{http://xxx.lanl.gov/abs/hep-th/0703022}{{\tt hep-th/0703022}}.

\bibitem{Pakman:2007hn}
A.~Pakman and A.~Sever, {\it Exact n=4 correlators of ads(3)/cft(2)},  {\em
  Phys. Lett.} {\bf B652} (2007) 60--62,
  [\href{http://xxx.lanl.gov/abs/arXiv:0704.3040 [hep-th]}{{\tt arXiv:0704.3040
  [hep-th]}}].

\bibitem{Jevicki:1998bm}
A.~Jevicki, M.~Mihailescu, and S.~Ramgoolam, {\it Gravity from cft on s**n(x):
  Symmetries and interactions},  {\em Nucl. Phys.} {\bf B577} (2000) 47--72,
  [\href{http://xxx.lanl.gov/abs/hep-th/9907144}{{\tt hep-th/9907144}}].

\bibitem{Lunin:2000yv}
O.~Lunin and S.~D. Mathur, {\it Correlation functions for m(n)/s(n) orbifolds},
   {\em Commun. Math. Phys.} {\bf 219} (2001) 399--442,
  [\href{http://xxx.lanl.gov/abs/hep-th/0006196}{{\tt hep-th/0006196}}].

\bibitem{Lunin:2001pw}
O.~Lunin and S.~D. Mathur, {\it Three-point functions for m(n)/s(n) orbifolds
  with n = 4 supersymmetry},  {\em Commun. Math. Phys.} {\bf 227} (2002)
  385--419, [\href{http://xxx.lanl.gov/abs/hep-th/0103169}{{\tt
  hep-th/0103169}}].

\bibitem{Mihailescu:1999cj}
M.~Mihailescu, {\it Correlation functions for chiral primaries in d = 6
  supergravity on ads(3) x s(3)},  {\em JHEP} {\bf 02} (2000) 007,
  [\href{http://xxx.lanl.gov/abs/hep-th/9910111}{{\tt hep-th/9910111}}].

\bibitem{Arutyunov:2000by}
G.~Arutyunov, A.~Pankiewicz, and S.~Theisen, {\it Cubic couplings in d = 6 n =
  4b supergravity on ads(3) x s(3)},  {\em Phys. Rev.} {\bf D63} (2001) 044024,
  [\href{http://xxx.lanl.gov/abs/hep-th/0007061}{{\tt hep-th/0007061}}].

\bibitem{Kanitscheider:2006zf}
I.~Kanitscheider, K.~Skenderis, and M.~Taylor, {\it Holographic anatomy of
  fuzzballs},  \href{http://xxx.lanl.gov/abs/hep-th/0611171}{{\tt
  hep-th/0611171}}.

\bibitem{Taylor:2007hs}
M.~Taylor, {\it Matching of correlators in $ads_3/cft_2$},
  \href{http://xxx.lanl.gov/abs/arXiv:0709.1838 [hep-th]}{{\tt arXiv:0709.1838
  [hep-th]}}.

\bibitem{Dijkgraaf:1998gf}
R.~Dijkgraaf, {\it Instanton strings and hyperkaehler geometry},  {\em Nucl.
  Phys.} {\bf B543} (1999) 545--571,
  [\href{http://xxx.lanl.gov/abs/hep-th/9810210}{{\tt hep-th/9810210}}].

\bibitem{Larsen:1999uk}
F.~Larsen and E.~J. Martinec, {\it U(1) charges and moduli in the d1-d5
  system},  {\em JHEP} {\bf 06} (1999) 019,
  [\href{http://xxx.lanl.gov/abs/hep-th/9905064}{{\tt hep-th/9905064}}].

\bibitem{Maldacena:1998bw}
J.~M. Maldacena and A.~Strominger, {\it Ads(3) black holes and a stringy
  exclusion principle},  {\em JHEP} {\bf 12} (1998) 005,
  [\href{http://xxx.lanl.gov/abs/hep-th/9804085}{{\tt hep-th/9804085}}].

\bibitem{Giveon:1998ns}
A.~Giveon, D.~Kutasov, and N.~Seiberg, {\it Comments on string theory on
  ads(3)},  {\em Adv. Theor. Math. Phys.} {\bf 2} (1998) 733--780,
  [\href{http://xxx.lanl.gov/abs/hep-th/9806194}{{\tt hep-th/9806194}}].

\bibitem{Kutasov:1999xu}
D.~Kutasov and N.~Seiberg, {\it More comments on string theory on ads(3)},
  {\em JHEP} {\bf 04} (1999) 008,
  [\href{http://xxx.lanl.gov/abs/hep-th/9903219}{{\tt hep-th/9903219}}].

\bibitem{deBoer:1998pp}
J.~de~Boer, H.~Ooguri, H.~Robins, and J.~Tannenhauser, {\it String theory on
  ads(3)},  {\em JHEP} {\bf 12} (1998) 026,
  [\href{http://xxx.lanl.gov/abs/hep-th/9812046}{{\tt hep-th/9812046}}].

\bibitem{Giveon:2003ku}
A.~Giveon and A.~Pakman, {\it More on superstrings in ads(3) x n},  {\em JHEP}
  {\bf 03} (2003) 056, [\href{http://xxx.lanl.gov/abs/hep-th/0302217}{{\tt
  hep-th/0302217}}].

\bibitem{Hikida:2000ry}
Y.~Hikida, K.~Hosomichi, and Y.~Sugawara, {\it String theory on ads(3) as
  discrete light-cone liouville theory},  {\em Nucl. Phys.} {\bf B589} (2000)
  134--166, [\href{http://xxx.lanl.gov/abs/hep-th/0005065}{{\tt
  hep-th/0005065}}].

\bibitem{Argurio:2000tb}
R.~Argurio, A.~Giveon, and A.~Shomer, {\it Superstrings on ads(3) and symmetric
  products},  {\em JHEP} {\bf 12} (2000) 003,
  [\href{http://xxx.lanl.gov/abs/hep-th/0009242}{{\tt hep-th/0009242}}].

\bibitem{Maldacena:2000hw}
J.~M. Maldacena and H.~Ooguri, {\it Strings in ads(3) and sl(2,r) wzw model.
  i},  {\em J. Math. Phys.} {\bf 42} (2001) 2929--2960,
  [\href{http://xxx.lanl.gov/abs/hep-th/0001053}{{\tt hep-th/0001053}}].

\bibitem{Maldacena:2000kv}
J.~M. Maldacena, H.~Ooguri, and J.~Son, {\it Strings in ads(3) and the sl(2,r)
  wzw model. ii: Euclidean black hole},  {\em J. Math. Phys.} {\bf 42} (2001)
  2961--2977, [\href{http://xxx.lanl.gov/abs/hep-th/0005183}{{\tt
  hep-th/0005183}}].

\bibitem{Maldacena:2001km}
J.~M. Maldacena and H.~Ooguri, {\it Strings in ads(3) and the sl(2,r) wzw
  model. iii: Correlation functions},  {\em Phys. Rev.} {\bf D65} (2002)
  106006, [\href{http://xxx.lanl.gov/abs/hep-th/0111180}{{\tt
  hep-th/0111180}}].

\bibitem{Seiberg:1999xz}
N.~Seiberg and E.~Witten, {\it The d1/d5 system and singular cft},  {\em JHEP}
  {\bf 04} (1999) 017, [\href{http://xxx.lanl.gov/abs/hep-th/9903224}{{\tt
  hep-th/9903224}}].

\bibitem{Raju:2007uj}
S.~Raju, {\it Counting giant gravitons in $ads_3$},
  \href{http://xxx.lanl.gov/abs/arXiv:0709.1171 [hep-th]}{{\tt arXiv:0709.1171
  [hep-th]}}.

\bibitem{Rastelli:2005ph}
L.~Rastelli and M.~Wijnholt, {\it Minimal ads(3)},
  \href{http://xxx.lanl.gov/abs/hep-th/0507037}{{\tt hep-th/0507037}}.

\bibitem{Martinec:1991ht}
E.~J. Martinec, G.~W. Moore, and N.~Seiberg, {\it Boundary operators in 2-d
  gravity},  {\em Phys. Lett.} {\bf B263} (1991) 190--194.

\bibitem{Vafa:1994tf}
C.~Vafa and E.~Witten, {\it A strong coupling test of s duality},  {\em Nucl.
  Phys.} {\bf B431} (1994) 3--77,
  [\href{http://xxx.lanl.gov/abs/hep-th/9408074}{{\tt hep-th/9408074}}].

\bibitem{D'Hoker:1999ea}
E.~D'Hoker, D.~Z. Freedman, S.~D. Mathur, A.~Matusis, and L.~Rastelli, {\it
  Extremal correlators in the ads/cft correspondence},
  \href{http://xxx.lanl.gov/abs/hep-th/9908160}{{\tt hep-th/9908160}}.

\bibitem{Fuchs:1988gm}
J.~Fuchs, {\it More on the super wzw theory},  {\em Nucl. Phys.} {\bf B318}
  (1989) 631.

\bibitem{Zamolodchikov:1986bd}
A.~B. Zamolodchikov and V.~A. Fateev, {\it Operator algebra and correlation
  functions in the two- dimensional wess-zumino su(2) x su(2) chiral model},
  {\em Sov. J. Nucl. Phys.} {\bf 43} (1986) 657--664.

\bibitem{Gepner:1986wi}
D.~Gepner and E.~Witten, {\it String theory on group manifolds},  {\em Nucl.
  Phys.} {\bf B278} (1986) 493.

\bibitem{Kutasov:1998zh}
D.~Kutasov, F.~Larsen, and R.~G. Leigh, {\it String theory in magnetic monopole
  backgrounds},  {\em Nucl. Phys.} {\bf B550} (1999) 183--213,
  [\href{http://xxx.lanl.gov/abs/hep-th/9812027}{{\tt hep-th/9812027}}].

\bibitem{Friedan:1985ge}
D.~Friedan, E.~J. Martinec, and S.~H. Shenker, {\it Conformal invariance,
  supersymmetry and string theory},  {\em Nucl. Phys.} {\bf B271} (1986) 93.

\bibitem{Feigin:1997ha}
B.~L. Feigin, A.~M. Semikhatov, and I.~Y. Tipunin, {\it Equivalence between
  chain categories of representations of affine sl(2) and n = 2 superconformal
  algebras},  {\em J. Math. Phys.} {\bf 39} (1998) 3865--3905,
  [\href{http://xxx.lanl.gov/abs/hep-th/9701043}{{\tt hep-th/9701043}}].

\bibitem{Maldacena:1998uz}
J.~M. Maldacena, J.~Michelson, and A.~Strominger, {\it Anti-de sitter
  fragmentation},  {\em JHEP} {\bf 02} (1999) 011,
  [\href{http://xxx.lanl.gov/abs/hep-th/9812073}{{\tt hep-th/9812073}}].

\bibitem{Israel:2003ry}
D.~Israel, C.~Kounnas, and M.~P. Petropoulos, {\it Superstrings on ns5
  backgrounds, deformed ads(3) and holography},  {\em JHEP} {\bf 10} (2003)
  028, [\href{http://xxx.lanl.gov/abs/hep-th/0306053}{{\tt hep-th/0306053}}].

\bibitem{Feigin:1998sw}
B.~L. Feigin, A.~M. Semikhatov, V.~A. Sirota, and I.~Y. Tipunin, {\it
  Resolutions and characters of irreducible representations of the n = 2
  superconformal algebra},  {\em Nucl. Phys.} {\bf B536} (1998) 617--656,
  [\href{http://xxx.lanl.gov/abs/hep-th/9805179}{{\tt hep-th/9805179}}].

\bibitem{Pakman:2003kh}
A.~Pakman, {\it Brst quantization of string theory in ads(3)},  {\em JHEP} {\bf
  06} (2003) 053, [\href{http://xxx.lanl.gov/abs/hep-th/0304230}{{\tt
  hep-th/0304230}}].

\bibitem{Aldazabal:1992ae}
G.~Aldazabal, M.~Bonini, and J.~M. Maldacena, {\it Factorization and discrete
  states in c = 1 superliouville theory},  {\em Int. J. Mod. Phys.} {\bf A9}
  (1994) 3969--3988, [\href{http://xxx.lanl.gov/abs/hep-th/9209010}{{\tt
  hep-th/9209010}}].

\bibitem{Witten:1991zd}
E.~Witten, {\it Ground ring of two-dimensional string theory},  {\em Nucl.
  Phys.} {\bf B373} (1992) 187--213,
  [\href{http://xxx.lanl.gov/abs/hep-th/9108004}{{\tt hep-th/9108004}}].

\bibitem{Kostelecky:1986xg}
V.~A. Kostelecky, O.~Lechtenfeld, W.~Lerche, S.~Samuel, and S.~Watamura, {\it
  Conformal techniques, bosonization and tree level string amplitudes},  {\em
  Nucl. Phys.} {\bf B288} (1987) 173.

\bibitem{Pakman:2003cu}
A.~Pakman, {\it Unitarity of supersymmetric sl(2,r)/u(1) and no-ghost theorem
  for fermionic strings in ads(3) x n},  {\em JHEP} {\bf 01} (2003) 077,
  [\href{http://xxx.lanl.gov/abs/hep-th/0301110}{{\tt hep-th/0301110}}].

\bibitem{Dotsenko:1992mg}
V.~S. Dotsenko, {\it Remarks on the physical states and the chiral algebra of
  $2-d$ gravity coupled to $c \leq 1$ matter},  {\em Theor. Math. Phys.} {\bf
  92} (1992) 938--951, [\href{http://xxx.lanl.gov/abs/hep-th/9201077}{{\tt
  hep-th/9201077}}].

\bibitem{Kitazawa:1987za}
Y.~Kitazawa {\em et~al.}, {\it Operator product expansion coefficients in n=1
  superconformal theory and slightly relevant perturbation},  {\em Nucl. Phys.}
  {\bf B306} (1988) 425.

\bibitem{AlvarezGaume:1991bj}
L.~Alvarez-Gaume and P.~Zaugg, {\it Structure constants in the n=1
  superoperator algebra},  {\em Ann. Phys.} {\bf 215} (1992) 171--230,
  [\href{http://xxx.lanl.gov/abs/hep-th/9109050}{{\tt hep-th/9109050}}].

\bibitem{Cappelli:1986hf}
A.~Cappelli, C.~Itzykson, and J.~B. Zuber, {\it Modular invariant partition
  functions in two-dimensions},  {\em Nucl. Phys.} {\bf B280} (1987) 445--465.

\bibitem{Cappelli:1987xt}
A.~Cappelli, C.~Itzykson, and J.~B. Zuber, {\it The ade classification of
  minimal and a1(1) conformal invariant theories},  {\em Commun. Math. Phys.}
  {\bf 113} (1987) 1.

\bibitem{Teschner:1997ft}
J.~Teschner, {\it On structure constants and fusion rules in the sl(2,c)/su(2)
  wznw model},  {\em Nucl. Phys.} {\bf B546} (1999) 390--422,
  [\href{http://xxx.lanl.gov/abs/hep-th/9712256}{{\tt hep-th/9712256}}].

\bibitem{Teschner:1999ug}
J.~Teschner, {\it Operator product expansion and factorization in the h-3+ wznw
  model},  {\em Nucl. Phys.} {\bf B571} (2000) 555--582,
  [\href{http://xxx.lanl.gov/abs/hep-th/9906215}{{\tt hep-th/9906215}}].

\bibitem{Becker:1993at}
K.~Becker and M.~Becker, {\it Interactions in the sl(2,ir) / u(1) black hole
  background},  {\em Nucl. Phys.} {\bf B418} (1994) 206--230,
  [\href{http://xxx.lanl.gov/abs/hep-th/9310046}{{\tt hep-th/9310046}}].

\bibitem{Giribet:1999ft}
G.~Giribet and C.~Nunez, {\it Interacting strings on ads(3)},  {\em JHEP} {\bf
  11} (1999) 031, [\href{http://xxx.lanl.gov/abs/hep-th/9909149}{{\tt
  hep-th/9909149}}].

\bibitem{Ishibashi:2000fn}
N.~Ishibashi, K.~Okuyama, and Y.~Satoh, {\it Path integral approach to string
  theory on ads(3)},  {\em Nucl. Phys.} {\bf B588} (2000) 149--177,
  [\href{http://xxx.lanl.gov/abs/hep-th/0005152}{{\tt hep-th/0005152}}].

\bibitem{Hosomichi:2000bm}
K.~Hosomichi, K.~Okuyama, and Y.~Satoh, {\it Free field approach to string
  theory on ads(3)},  {\em Nucl. Phys.} {\bf B598} (2001) 451--466,
  [\href{http://xxx.lanl.gov/abs/hep-th/0009107}{{\tt hep-th/0009107}}].

\bibitem{Hosomichi:2001fm}
K.~Hosomichi and Y.~Satoh, {\it Operator product expansion in string theory on
  ads(3)},  {\em Mod. Phys. Lett.} {\bf A17} (2002) 683--693,
  [\href{http://xxx.lanl.gov/abs/hep-th/0105283}{{\tt hep-th/0105283}}].

\bibitem{Satoh:2001bi}
Y.~Satoh, {\it Three-point functions and operator product expansion in the
  sl(2) conformal field theory},  {\em Nucl. Phys.} {\bf B629} (2002) 188--208,
  [\href{http://xxx.lanl.gov/abs/hep-th/0109059}{{\tt hep-th/0109059}}].

\bibitem{FZZ}
V.~Fateev, A.~B. Zamolodchikov, and A.~B. Zamolodchikov, {\it unpublished}, .

\bibitem{Giribet:2000fy}
G.~Giribet and C.~Nunez, {\it Aspects of the free field description of string
  theory on ads(3)},  {\em JHEP} {\bf 06} (2000) 033,
  [\href{http://xxx.lanl.gov/abs/hep-th/0006070}{{\tt hep-th/0006070}}].

\bibitem{Giribet:2001ft}
G.~Giribet and C.~Nunez, {\it Correlators in ads(3) string theory},  {\em JHEP}
  {\bf 06} (2001) 010, [\href{http://xxx.lanl.gov/abs/hep-th/0105200}{{\tt
  hep-th/0105200}}].

\bibitem{Hofman:2004ny}
D.~M. Hofman and C.~A. Nunez, {\it Free field realization of superstring theory
  on ads(3)},  {\em JHEP} {\bf 07} (2004) 019,
  [\href{http://xxx.lanl.gov/abs/hep-th/0404214}{{\tt hep-th/0404214}}].

\bibitem{Giribet:2004zd}
G.~E. Giribet and D.~E. Lopez-Fogliani, {\it Remarks on free field realization
  of sl(2,r)k/u(1) x u(1) wznw model},  {\em JHEP} {\bf 06} (2004) 026,
  [\href{http://xxx.lanl.gov/abs/hep-th/0404231}{{\tt hep-th/0404231}}].

\bibitem{Giribet:2005ix}
G.~Giribet and Y.~Nakayama, {\it The stoyanovsky-ribault-teschner map and
  string scattering amplitudes},  {\em Int. J. Mod. Phys.} {\bf A21} (2006)
  4003--4034, [\href{http://xxx.lanl.gov/abs/hep-th/0505203}{{\tt
  hep-th/0505203}}].

\bibitem{Ribault:2005ms}
S.~Ribault, {\it Knizhnik-zamolodchikov equations and spectral flow in ads(3)
  string theory},  {\em JHEP} {\bf 09} (2005) 045,
  [\href{http://xxx.lanl.gov/abs/hep-th/0507114}{{\tt hep-th/0507114}}].

\bibitem{Giribet:2005mc}
G.~Giribet, {\it On spectral flow symmetry and knizhnik-zamolodchikov
  equation},  {\em Phys. Lett.} {\bf B628} (2005) 148--156,
  [\href{http://xxx.lanl.gov/abs/hep-th/0508019}{{\tt hep-th/0508019}}].

\bibitem{Minces:2005nb}
P.~Minces, C.~Nunez, and E.~Herscovich, {\it Winding strings in ads(3)},  {\em
  JHEP} {\bf 06} (2006) 047,
  [\href{http://xxx.lanl.gov/abs/hep-th/0512196}{{\tt hep-th/0512196}}].

\bibitem{Minces:2007td}
P.~Minces and C.~Nunez, {\it Four point functions in the sl(2,r) wzw model},
  {\em Phys. Lett.} {\bf B647} (2007) 500--508,
  [\href{http://xxx.lanl.gov/abs/hep-th/0701293}{{\tt hep-th/0701293}}].

\bibitem{Iguri:2007af}
S.~Iguri and C.~Nunez, {\it Coulomb integrals for the sl(2,r) wzw model},
  \href{http://xxx.lanl.gov/abs/arXiv:0705.4461 [hep-th]}{{\tt arXiv:0705.4461
  [hep-th]}}.

\bibitem{Aharony:2003vk}
O.~Aharony, B.~Fiol, D.~Kutasov, and D.~A. Sahakyan, {\it Little string theory
  and heterotic/type ii duality},  {\em Nucl. Phys.} {\bf B679} (2004) 3--65,
  [\href{http://xxx.lanl.gov/abs/hep-th/0310197}{{\tt hep-th/0310197}}].

\bibitem{Christe:1986cy}
P.~Christe and R.~Flume, {\it The four point correlations of all primary
  operators of the d = 2 conformally invariant su(2) sigma model with wess-
  zumino term},  {\em Nucl. Phys.} {\bf B282} (1987) 466.

\bibitem{Dotsenko:1990zb}
V.~S. Dotsenko, {\it Solving the su(2) conformal field theory with the wakimoto
  free field representation},  {\em Nucl. Phys.} {\bf B358} (1991) 547--570.

\bibitem{Zamolodchikov:1995aa}
A.~B. Zamolodchikov and A.~B. Zamolodchikov, {\it Structure constants and
  conformal bootstrap in liouville field theory},  {\em Nucl. Phys.} {\bf B477}
  (1996) 577--605, [\href{http://xxx.lanl.gov/abs/hep-th/9506136}{{\tt
  hep-th/9506136}}].

\bibitem{Kostov:2005av}
I.~K. Kostov and V.~B. Petkova, {\it Non-rational 2d quantum gravity. i: World
  sheet cft},  \href{http://xxx.lanl.gov/abs/hep-th/0512346}{{\tt
  hep-th/0512346}}.

\end{thebibliography}\endgroup

\bibliographystyle{JHEP}

\end{document}